\documentclass[onecolumn,amsmath,showpacs,11pt]{revtex4}
\usepackage{graphicx}
\usepackage{dcolumn}
\usepackage{bm}
\begin{document}
\newcommand{\hs}{\hspace*{0.5cm}}
\newcommand{\vs}{\vspace*{0.5cm}}
\newcommand{\be}{\begin{equation}}
\newcommand{\ee}{\end{equation}}
\newcommand{\bea}{\begin{eqnarray}}
\newcommand{\eea}{\end{eqnarray}}
\newcommand{\ben}{\begin{enumerate}}
\newcommand{\een}{\end{enumerate}}
\newcommand{\bde}{\begin{widetext}}
\newcommand{\ede}{\end{widetext}}
\newcommand{\nn}{\nonumber}
\newcommand{\crn}{\nonumber \\}
\newcommand{\Tr}{\mathrm{Tr}}
\newcommand{\non}{\nonumber}
\newcommand{\noi}{\noindent}
\newcommand{\al}{\alpha}
\newcommand{\la}{\lambda}
\newcommand{\bet}{\beta}
\newcommand{\ga}{\gamma}
\newcommand{\va}{\varphi}
\newcommand{\om}{\omega}
\newcommand{\pa}{\partial}
\newcommand{\+}{\dagger}
\newcommand{\fr}{\frac}
\newcommand{\bc}{\begin{center}}
\newcommand{\ec}{\end{center}}
\newcommand{\Ga}{\Gamma}
\newcommand{\de}{\delta}
\newcommand{\De}{\Delta}
\newcommand{\ep}{\epsilon}
\newcommand{\varep}{\varepsilon}
\newcommand{\ka}{\kappa}
\newcommand{\La}{\Lambda}
\newcommand{\si}{\sigma}
\newcommand{\Si}{\Sigma}
\newcommand{\ta}{\tau}
\newcommand{\up}{\upsilon}
\newcommand{\Up}{\Upsilon}
\newcommand{\ze}{\zeta}
\newcommand{\ps}{\psi}
\newcommand{\Ps}{\Psi}
\newcommand{\ph}{\phi}
\newcommand{\vph}{\varphi}
\newcommand{\Ph}{\Phi}
\newcommand{\Om}{\Omega}

\title{Inflation and leptogenesis in the 3-3-1-1 model}

\author{D. T. Huong}
\email {dthuong@iop.vast.ac.vn} \affiliation{Institute of Physics, Vietnam Academy of Science and Technology, 10 Dao Tan, Ba Dinh, Hanoi, Vietnam}
\author{P. V. Dong}
\email {pvdong@iop.vast.ac.vn} \affiliation{Institute of Physics, Vietnam Academy of Science and Technology, 10 Dao Tan, Ba Dinh, Hanoi, Vietnam}

\author{C. S. Kim}
\email {cskim@yonsei.ac.kr} \affiliation{Department of Physics and IPAP, Yonsei University, Seoul 120-479, Korea}
\author{N. T. Thuy}
\email {ntthuy@iop.vast.ac.vn} \affiliation{Department of Physics and IPAP, Yonsei University, Seoul 120-479, Korea}

\date{\today}

\begin{abstract}
We consider the $SU(3)_C \otimes SU(3)_L \otimes U(1)_X \otimes U(1)_N $ (3-3-1-1) model at the GUT scale with implication for inflation and leptogenesis. The mass spectra of the neutral Higgs bosons and neutral gauge bosons are reconsidered when the scale of the 3-3-1-1 breaking is much larger than that of the ordinary $SU(3)_C \otimes SU(3)_L \otimes U(1)_X $ (3-3-1) breaking. We investigate how the 3-3-1-1 model generates an inflation by identifying the scalar field that spontaneously breaks the $U(1)_N$ symmetry to inflaton as well as including radiative corrections for the inflaton potential. We
figure out the parameter spaces appeared in the inflaton potential that satisfy the conditions for an inflation model and obtain the inflaton mass an order of $10^{13}$ GeV. The inflaton can dominantly decay into a pair of light Higgs bosons or a pair of heavy Majorana neutrinos which lead, respectively, to a reheating temperature of $10^9$ GeV order appropriate to a thermal leptogenesis scenario or to a reduced reheating temperature corresponding to a non-thermal leptogenesis scenario. We calculate the lepton asymmetry which yields baryon asymmetry successfully for both the thermal and non-thermal cases.
\end{abstract}


\maketitle
\newpage

\section{\label{intro}Introduction}

Cosmological inflation is a popular postulate for the early universe. It can solve
the difficulties of the hot Big Bang theory and provide the predictions for quantum fluctuations
in the inflating background. In order to recover the conventions of the hot Big Bang theory and
to know how the universe is reheated, we must understand what is the inflaton field, and how it is
connected to particle physics. These problems were first investigated with the chaotic inflation scenario
by Linde \cite{c1}. According to this scenario, the inflation may begin  even there was no thermal
equilibrium in the early universe. It can occur in a theory with a very simple potential
such as $V(\phi)\propto \phi^2.$ There is no limit to the theory with a polynomial
potential: Chaotic inflation occurs in any theory where the potential has a sufficiently flat region \cite{c1}.
On the other hand, the recent measurements of $B$ modes by BICEP2 collaboration \cite{bicep2} have yielded very interesting
results, which could be the direct measurements of quantum gravitation excitations from the early universe.
The ratio of the tensor and scalar is measured as $0.16^{+ 0.06}_{-0.05}$.
Combined with the Planck and WMAP measurements suggests that the inflation model must be a larger field model.
Hence, the inflationary scenario does not work on the framework of the Standard Model (SM) without a non-minimal coupling to gravity.

Furthermore, what is the origin of matter-antimatter asymmetry in the universe? The neutrino experiments such as Super-Kamiokande \cite{c2}, KamLAND \cite{c2a} and
SNO \cite{c3a} have confirmed that the neutrinos have small masses and large flavor mixing. According to the Planck mission team, and based
on the standard model of cosmology, there exits dark matter (DM) which lies beyond the SM. All the experiments call for extensions beyond the SM.
One way to extend the SM is to expand the gauge symmetry group. There exists a simple
extension of the SM gauge group to $SU(3)_C \otimes SU(3)_L \otimes U(1)_X$, the so-called 3-3-1
models. These models can explain the following issues \cite{331};
\begin{itemize}
\item Why the electric charges are quantized?
\item Why there are only three observed families of fermions?
\item Why top quark is oddly heavy?
\item Why the strong CP nonconservation is disappeared?
\item 3-3-1 models can provide the neutrino small masses as well as candidates for the DM
\cite{331DM, c4, c5}.
\end{itemize}

 There have recently emerged an extension of the 3-3-1 models,
 based on the $SU(3)_C \otimes SU(3)_L \otimes U(1)_X \otimes U(1)_N$ (3-3-1-1) gauge group, which not only contains all the good features of the 3-3-1 models as mentioned \cite{c4,c5}, but also has the following advantages;
 \begin{itemize}
 \item The $B-L$ number is naturally gauged by
 combination of the $SU(3)_L$ and $U(1)_N$ charges. It leads to an unification of the electroweak and $B-L$ interactions.
 \item The right-handed neutrinos appear in the model as fundamental fermions that solve the
  small masses of neutrinos through a type I seesaw mechanism.
 \item There exists a W-parity symmetry as a
($Z_2$) remnant subgroup of the gauge symmetry. Almost all the new particles have wrong lepton
 numbers transforming as odd fields under W-parity. The lightest wrong lepton number particle is
 identified to the DM. Because of W-parity conservation, the model can work better under experimental
 constraints than the 3-3-1 models.
\end{itemize}

 Other highlights of the 3-3-1-1 model is that the energy scale of the symmetry breaking
 $U(1)_N$, which can happen at a very high scale like the GUT one \cite{c4}. The inflationary
scenario can be linked to $U(1)_N$ breaking and driven by the Higgs $\phi$ potential. Due to the local
gauge $U(1)_N$ symmetry, a radiative correction to the inflaton potential can arise from the coupling
of inflaton with the $U(1)_N$ gauge boson $(Z_2)$. There exist the couplings of inflaton $\phi$ with
right handed neutrinos and Higgs triplets, which also contribute to the
inflation potential. We would like to stress that the well-known advantages of a spontaneously
broken gauge $U(1)_N$ symmetry include a seesaw mechanism for the neutrino physics \cite{c4}. The
presence of the right-handed neutrinos that directly interact to the inflaton may be compatible
with the leptogenesis scenario. The aim of this work is to show that the chaotic inflationary
scenario can be driven by the singlet Higgs $\phi$ potential. We also focus on the
leptogenesis happened after the inflation through the couplings of the right-handed neutrinos.

Our paper is organized as follows: In section \ref{intr}, we briefly review the 3-3-1-1 model and
specially concentrate on the Higgs and gauge boson spectra in the large $\Lambda$ limit. In section
\ref{inf}, we present the inflation model by assuming the singlet Higgs $\phi$ as an inflation field.
The leptogenesis related to the matter-antimatter asymmetry of the universe and neutrino properties
is studied in section \ref{lept}. Finally we summarize our works in section \ref{conc}.

\section{\label{intr}Brief description of the 3-3-1-1 model}

The fermion content of the 3-3-1-1 model is given as \cite{c4,c5} \bea \psi_{aL} &=&
\left(\begin{array}{c}
               \nu_{aL}\\ e_{aL}\\ (N_{aR})^c
\end{array}\right) \sim (1,3, -1/3,-2/3),\\
\nu_{aR}&\sim&(1,1,0,-1),\hs e_{aR} \sim (1,1, -1,-1),
\\
Q_{\al L}&=&\left(\begin{array}{c}
  d_{\al L}\\  -u_{\al L}\\  D_{\al L}
\end{array}\right)\sim (3,3^*,0,0),\hs Q_{3L}=\left(\begin{array}{c} u_{3L}\\  d_{3L}\\ U_L \end{array}\right)\sim
 \left(3,3,1/3,2/3\right),\\ u_{a
R}&\sim&\left(3,1,2/3,1/3\right),\hs d_{a R} \sim \left(3,1,-1/3,1/3\right),\\ U_{R}&\sim&
\left(3,1,2/3,4/3\right),\hs D_{\al R} \sim \left(3,1,-1/3,-2/3\right),\eea where the quantum
numbers in the parentheses are defined upon the gauge symmetries $(SU(3)_C,$ $\\\ SU(3)_L,\ U(1)_X$
$,\ U(1)_N)$, respectively. The family indices are $a=1,2,3$ and $\al=1,2$. $N_R$, $U$ and $D$ are
the exotic fermions, which have incorrect lepton numbers. The other fermions have ordinary lepton
numbers. Note that the neutral fermions $N_{R}$ are truly sterile since they do not have any gauge
interaction, which contradicts to the $\nu_R$ ones as usually considered.

To break the gauge symmetry, one uses the following scalar multiplets \cite{c4}:
\bea \rho &=& \left(\begin{array}{c}
\rho^+_1\\
\rho^0_2\\
\rho^+_3\end{array}\right)\sim (1,3,2/3,1/3), \hs \hs  \eta =  \left(\begin{array}{c}
\eta^0_1\\
\eta^-_2\\
\eta^0_3\end{array}\right)\sim (1,3,-1/3,1/3),\label{vev1}\\
 \chi &=& \left(\begin{array}{c}
\chi^0_1\\
\chi^-_2\\
\chi^0_3\end{array}\right)\sim (1,3,-1/3,-2/3), \hs \hs  \phi \sim (1,1,0,2),\eea with the VEVs
that conserve electric charge and $R$-parity being respectively given by \bea \langle \rho \rangle =
\fr{1}{\sqrt{2}}(0,v,0)^T,\hs \langle \eta \rangle =\fr{1}{\sqrt{2}}(u,0,0)^T,\hs \langle \chi
\rangle =\fr{1}{\sqrt{2}}(0,0,\om)^T,\hs \langle \phi\rangle =\fr{1}{\sqrt{2}}\La.\label{vevss}
\eea
The pattern of the symmetry breaking of the model is given by the following scheme  \bea \mbox{
3-3-1-1}&\stackrel{\langle\chi\rangle \langle \rho\rangle\langle \eta \rangle}{\longmapsto}&
\mbox{SU(3)}_{C} \ \otimes \ \mbox{U(1)}_{Q}\otimes \mbox{U(1)}_{B-L} \stackrel{\langle\phi\rangle
}{\longmapsto} \mbox{SU(3)}_{C} \ \otimes \ \mbox{U(1)}_{Q} \ \otimes \mbox{P},
\label{breaksusy331tou1} \eea where the electric charge $Q$, $B-L$ and matter parity $P$ take the forms,
\bea Q=T_3-\fr{1}{\sqrt{3}}T_8+X,\hs
B-L=-\fr{2}{\sqrt{3}}T_8+N, \hs P=(-1)^{3(B-L)}.\eea Here, $T_i\ (i=1,2,3,...,8)$, $X$ and $N$ are the $SU(3)_L$, $U(1)_X$ and $U(1)_N$ charges, respectively.

The Lagrangian of the 3-3-1-1 model is given by \cite{c4}:
\bea \mathcal{L}&=&\sum_{\mathrm{fermion\
multiplets}}\bar{\Psi}i\ga^\mu D_\mu \Psi + \sum_{\mathrm{scalar\ multiplets}}(D^\mu \Phi)^\dagger
(D_\mu \Phi)\crn && -\fr{1}{4}G_{i\mu\nu}G_i^{\mu\nu} -\fr 1 4 A_{i\mu\nu}A_i^{\mu\nu}-\fr 1 4
B_{\mu\nu} B^{\mu\nu}-\fr{1}{4}C_{\mu\nu} C^{\mu\nu}\crn &&-V(\rho,\eta,\chi,\phi) +
\mathcal{L}_{\mathrm{Yukawa}},\eea
where the  Yukawa Lagrangian and scalar potential are obtained  \cite{c4,c5} as follows
\bea
\mathcal{L}_{\mathrm{Yukawa}}&=&h^e_{ab}\bar{\psi}_{aL}\rho e_{bR}
+h^\nu_{ab}\bar{\psi}_{aL}\eta\nu_{bR}+h'^\nu_{ab}\bar{\nu}^c_{aR}\nu_{bR}\phi +
h^U\bar{Q}_{3L}\chi U_R + h^D_{\al \beta}\bar{Q}_{\al L} \chi^* D_{\beta R}\crn
&&+ h^u_a \bar{Q}_{3L}\eta u_{aR}+h^d_a\bar{Q}_{3L}\rho d_{aR} + h^d_{\al a} \bar{Q}_{\al L}\eta^* d_{aR} +h^u_{\al a } \bar{Q}_{\al L}\rho^* u_{aR} +H.c,\\
V(\rho,\eta,\chi,\phi) &=& \mu^2_1\rho^\dagger \rho + \mu^2_2 \chi^\dagger \chi + \mu^2_3
\eta^\dagger \eta + \la_1 (\rho^\dagger \rho)^2 + \la_2 (\chi^\dagger \chi)^2 + \la_3 (\eta^\dagger
\eta)^2\crn &&+ \la_4 (\rho^\dagger \rho)(\chi^\dagger \chi) +\la_5 (\rho^\dagger
\rho)(\eta^\dagger \eta)+\la_6 (\chi^\dagger \chi)(\eta^\dagger \eta)\crn && +\la_7 (\rho^\dagger
\chi)(\chi^\dagger \rho) +\la_8 (\rho^\dagger \eta)(\eta^\dagger \rho)+\la_9 (\chi^\dagger
\eta)(\eta^\dagger \chi) + (f\epsilon^{mnp}\eta_m\rho_n\chi_p+H.c.) \crn && + \mu^2 \phi^\dagger
\phi + \la (\phi^\dagger \phi)^2 +\la_{10} (\phi^\dagger
\phi)(\rho^\dagger\rho)+\la_{11}(\phi^\dagger \phi)(\chi^\dagger \chi)+\la_{12}(\phi^\dagger
\phi)(\eta^\dagger \eta).
\label{scalar}
\eea
Because of the 3-3-1-1 gauge symmetry, the Yukawa
Lagangian and scalar potential as given take the standard forms which contain no lepton-number
violating interactions.

The fermion masses that result from the Yukawa Lagrangian have been presented in \cite{c4}. The phenomenology of the 3-3-1-1 model with the $\La$ scale of the $U(1)_N$ breaking comparable to the $\om$ scale of
the 3-3-1 symmetry breaking has been studied
in \cite{c5}. Below, we will compute the physical states and masses for the scalar and gauge sectors in the limit $\La \gg \om$, which is needed for our further analysis.

\subsection{\label{sclsec}Scalar sector}

In this part, we identify the physical particles in the scalar sector. We expand the neutral scalars around their VEVs \cite{c4}
such as
\bea
\rho &=& \left(\begin{array}{c}
\rho^+_1\\
\frac{1}{\sqrt{2}}(v+S_2+iA_2)\\
\rho^+_3\end{array}\right);
\eta = \left(\begin{array}{c}
\frac{1}{\sqrt{2}}(u+S_1+iA_1)\\
\eta^-_2\\
\frac{1}{\sqrt{2}}(S_3^\prime+iA_3^\prime)\end{array}\right);
 \chi = \left(\begin{array}{c}
\frac{1}{\sqrt{2}}(S_1^\prime+i A_1^\prime)\\
\chi^-_2\\
\frac{1}{\sqrt{2}}(\om+S_3+iA_3)\end{array}\right);\\
\phi &\sim& \frac{1}{\sqrt{2}}(\Lambda+S_4+i A_4).\eea

In the scalar sector, all scalar fields with W-parity even, $S_1, S_2, S_3, S_4$, mix via the mass matrix such as
\bea M_{S}^2= \left(\begin{array}{cccc}
 2\la_3 u^2-\fr{1}{\sqrt{2}}f\fr{v\om}{u} & \la_5 uv+\fr{1}{\sqrt{2}}f\om & \la_6 u\om+\fr{1}{\sqrt{2}}fv & \la_{12}u\La\\
\la_5 uv+\fr{1}{\sqrt{2}}f\om & 2\la_1 v^2-\fr{1}{\sqrt{2}}f\fr{u\om}{v} & \la_4 \om v+\fr{1}{\sqrt{2}}fu & \la_{10}v\La\\
\la_6 u\om+\fr{1}{\sqrt{2}}fv & \la_4 \om v+\fr{1}{\sqrt{2}}fu & 2\la_2 \om^2-\fr{1}{\sqrt{2}}f\fr{vu}{\om} &  \la_{11}\om\La\\
\la_{12}u\La & \la_{10}v\La & \la_{11}\om\La & 2\la\La^2
\end{array}\right)\label{neutral1}. \eea
 We assume that $ \La \gg \om \sim -f \gg u,v$
then the mass matrix given in
 Eq. (\ref{neutral1}) has form as

\bea
M_{S}^2 = \left (\begin{array}{cc}
C & B^T \\
B & A
\end{array} \right ), \label{neutral2}
\eea
 where
\be
A=2\la \La^2,
\ee
\be
B=\left(
  \begin{array}{ccc}
    \la_{12 }u\La &  \la_{10} v\La & \la_{11} \om\La \\
  \end{array}
\right),
\ee
\be
C=\left(
    \begin{array}{ccc}
      2 \la_3u^2 -\frac{fv\om}{\sqrt{2}u} & \la_5 u v +\frac{f\om}{ \sqrt{2}}
      & \la_6 u \om +\frac{fv}{\sqrt{2}}\\
      \la_5 u v +\frac{f\om}{ \sqrt{2}} & 2 \la_1 v^2-\frac{fu\om}{\sqrt{2}v}
      & \la_4 v\om + \frac{fu}{\sqrt{2}} \\
      \la_6 u \om +\frac{fv}{\sqrt{2}} & \la_4 v\om + \frac{fu}{\sqrt{2}}
      &2\la_2 \om^2 -\frac{fuv}{\sqrt{2}\om} \\
    \end{array}
  \right).
\ee

Since $(\La \gg -f,\om\gg u,v)$, we get $A\gg  B,C$. The matrix given in
(\ref{neutral2}) can be diagonalized by using block diagonalizing  method.
The approximately unitary matrix $U$,
\be U=
\left(
  \begin{array}{cc}
    1 & B^\+ A^{-1} \\
    -A^{-1} B & 1 \\
  \end{array}
\right)=
\left(
  \begin{array}{cccc}
    1 & 0 & 0 & \fr{u \la_{12}}{2 \la \La}\\
    0 & 1 & 0 & \fr{v \la_{10}}{2 \la \La}  \\
    0 & 0 & 1 & \fr{\om \la_{11}}{2 \la \La} \\
    -\fr{u \la_{12}}{2 \la \La}  & -\fr{v \la_{10}}{2 \la \La} &
      -\fr{\om \la_{11}}{2 \la \La}  & 1 \\
  \end{array}
\right), \ee transform $M_S^2$ into approximately block-diagonal form: \be U^\+ M_S^2 U \approx
\left(
  \begin{array}{cc}
    C-B^\+ A^{-1} B & 0 \\
    0 & A \\
  \end{array}
\right).
\ee
In the limit $(\La \gg -f,\om\gg u,v)$, $U\simeq I $ then $ H_3 \simeq S_4 $ gets mass
$m_{H_3}^2= 2 \la \La^2$.
$S_1,S_2,S_3$ are mixing with the mixing mass matrix obtained as

\be
C-B^\+ A^{-1} B=\left(
    \begin{array}{ccc}
       2 \la_3u^2 -\frac{fv\om}{\sqrt{2}u} - \fr{u^2 \la_{12}^2}{2 \la} &
       \la_5 u v +\frac{f\om}{ \sqrt{2}}  - \fr{u v \la_{10}\la_{12}}{2 \la}&
      \la_6 u \om +\frac{fv}{\sqrt{2}}  - \fr{ u\om \la_{11}\la_{12}}{2 \la} \\
      \la_5 u v +\frac{f\om}{ \sqrt{2}}  - \fr{u v \la_{10}\la_{12}}{2 \la}&
      2\la_1 v^2-\frac{fu\om}{\sqrt{2}v}- \fr{v^2 \la_{10}^2}{2 \la} &
      \la_4 v\om + \frac{fu}{\sqrt{2}} - \fr{ v\om \la_{10}\la_{11}}{2 \la}\\
      \la_6 u \om +\frac{fv}{\sqrt{2}}  - \fr{ u\om \la_{11}\la_{12}}{2 \la}&
      \la_4 v\om + \frac{fu}{\sqrt{2}} - \fr{ v\om \la_{10}\la_{11}}{2 \la} &
      2\la_2 \om^2 -\frac{fuv}{\sqrt{2}\om} - \fr{\om^2 \la_{11}^2}{2 \la}\\
    \end{array}
  \right).\label{neutral3}
\ee
At the leading order $(-f,\om \gg u,v)$, the mass matrix given in Eq. (\ref{neutral3} ) can be rewritten as
\be
\left(
  \begin{array}{ccc}
    -\frac{fv\om}{\sqrt{2}u}& \frac{f\om}{ \sqrt{2}} & 0 \\
    \frac{f\om}{ \sqrt{2}} & -\frac{fu\om}{\sqrt{2}v} & 0 \\
    0 & 0 & 2\la_2 \om^2 - \fr{\om^2 \la_{11}^2}{2 \la}\\
  \end{array}
\right).
\ee
The physical fields with respective masses can be written as
\bea
H&=&\frac{uS_1+vS_2}{\sqrt{u^2+v^2}}, \hs
m_{H}^2 = 0, \crn
H_1&=&\frac{-vS_1+uS_2}{\sqrt{u^2+v^2}}, \hs
m^2_{H_1}= -\frac{f(u^2+v^2)\om}{\sqrt{2}uv},\crn
H_2&=& S_3 ,\hs m^2_{H_2}= \fr { (4 \la \la_2-\la_{11}^2)\om^2}{2 \la}.
\eea
In the new basics, (${H,H_1, H_2}$), the squared mass matrix given in
(\ref{neutral3}) can be written as

\bea
\left (\begin{array}{cc}
C^\prime & B^{\prime T} \\
B^\prime & A^\prime
\end{array} \right ), \label{neutral4}
\eea
 where
\be
C^\prime=\fr{v^4 (4 \la \la_1 - \la_{10}^2) - u^4 (\la_{12}^2 - 4 \la\la_3)
    -2 u^2 v^2 (\la_{10} \la_{11} - 2 \la \la_5)}{2 (u^2 + v^2) \la},
\ee
\be
B^\prime=\left(
      \begin{array}{c}
        \fr{u v (v^2 (\la_{10} (-\la_{10} + \la_{11}) + \la (4 \la_1 - 2 \la_5))
        + u^2 (-\la_{10} \la_{11} + \la_{11}^2 -
         4 \la \la_3 + 2 \la \la_5))}{2 (u^2 + v^2) \la} \\
        \fr{2 \sqrt2 f u v \la - \om (v^2 (\la_{10} \la_{11} - 2 \la \la_4) +
         u^2 (\la_{11}\la_{12}- 2 \la \la_6))}{2 \sqrt{u^2 + v^2} \la} \\
      \end{array}
    \right),
\ee
\be
A^\prime=\left(
           \begin{array}{cc}
             \fr{-\sqrt2 f (u^2 + v^2)^2 \om \la + u^3 v^3 (-(\la_{10} - \la_{12})^2 +
              4 \la (\la_1 + \la_3 - \la_5))}{2 u v(u^2 + v^2) \la} &
              \fr{\sqrt2 f (u^2 - v^2) \la +  u v \om (-\la_{10} \la_{11} + \la_{11} \la_{12} +
              2 \la \la_4 - 2 \la \la_6)}{2  \sqrt{u^2 + v^2}  \la} \\
              \fr{\sqrt2 f (u^2 - v^2) \la +  u v \om (-\la_{10} \la_{11} + \la_{11} \la_{12} +
              2 \la \la_4 - 2 \la \la_6)}{2  \sqrt{u^2 + v^2}  \la} &
              -\fr{\sqrt2 f u v \la + \om^3 (\la_{11}^2 - 4 \la \la_2)}{2 \om \la} \\
           \end{array}
         \right).
\ee
Since $-f,\om \gg u,v$, we get $A^\prime \gg B^\prime, C^\prime$.
If we kept explicitly the
$\mathcal{O}(\frac{u,v}{\om})$, the $H_1,H_2,H$ Higgs bosons can gain mass by using block
diagonalizing method as \bea m_{H_1}^{\prime 2}&=& m_{H_1}^{ 2}+\mathcal{O}(\frac{u,v}{\om}), \crn
m_{H_2}^{\prime 2}&=& m_{H_2}^{ 2}+\mathcal{O}(\frac{u,v}{\om}), \crn
m_{H}^{\prime 2}&=&\fr{v^4 (4 \la \la_1 - \la_{10}^2) - u^4 (\la_{12}^2 - 4 \la\la_3)
    -2 u^2 v^2 (\la_{10} \la_{12} - 2 \la \la_5)}{2 (u^2 + v^2) \la}\crn
    &&+ \fr 1{2 \sqrt2 (u^2 + v^2) \la (\la_{11}^2 - 4 \la\la_2)}
    (m_0+m_1 \fr f {\om}+m_2 \fr {f^2} {\om^2}),
\eea
where
\bea
m_0&=&\sqrt2 (v^2 (\la_{10} \la_{11} - 2 \la \la_4) +
   u^2 (\la_{11} \la_{12} - 2 \la \la_6))^2,\crn
m_1&=&8 u v \la (v^2 (-\la_{10} \la_{11} +
      2 \la \la_4) + u^2 (-\la_{11}\la_{12}+ 2 \la \la_6)),\crn
m_2&=&8 \sqrt2 u^2 v^2 \la^2.
\eea

For the remaining fields in the pseudoscalar sector, the mass spectrum is similar to that of work
given in \cite{c5}. Let us give a brief result. \begin{itemize}

\item The pseudoscalar $A_4$ is massless and is identified to the Goldstone boson of $Z_N$.
\item  Two other fields are massless that are identified to the Goldstone bosons of Z and $Z^\prime$
\bea G_Z= \frac{-uA_1+vA_2}{\sqrt{u^2+v^2}}; \hs \hs
G_{Z^\prime}=\frac{-\om^{-1}(u^{-1}A_1+v^{-1}A_2)+(u^{-2}+v^{-2})A_3}
{\sqrt{(u^{-2}+v^{-2}+\om^{-2})(u^{-2}+v^{-2})}}. \eea \item One neutral complex Goldstone boson ,
$G_X=\frac{\om \chi_1
  -u \eta_3^*}{\sqrt{u^2+\om^2}}$, that is eaten by X gauge boson.
\item One neutral complex Higgs , namely $H^\prime
  =\frac{u \chi^*_1+\om \eta_3}{\sqrt{u^2+\om^2}}$ with the squared mass $m_{H^\prime}^2 =(\frac{1}{2}\la_9 -\frac{fv}{\sqrt{2}u\om})(u^2+\om^2)$.
\item One physical pseudoscalar ($A$) with mass \bea m_A^2 =-\frac{f}{\sqrt{2}}
\frac{u^2v^2+u^2\om^2+v^2\om^2}{uv\om}, \eea and the physical state respectively \bea
A=\frac{u^{-1}A_1+v^{-1}A_2+\om^{-1}A_3}{\sqrt{u^{-2}+v^{-2}+\om^{-2}}}.\eea

For charged scalars, the mass spectrum is seminar to that of work given in \cite{c4}. \be
H_4^-=\fr{v \chi_2^- + \om \rho_3^-}{\sqrt{v^2+\om^2}}, \hs H_5^-=\fr{v \eta_2^- + u
\rho_1^-}{\sqrt{u^2+v^2}}, \ee with respective masses \be m_{H_4}^2=\left(\fr 1 2 \la_7- \fr {f u}
{\sqrt 2 v \om}\right) (v^2+\om^2), \hs m_{H_5}^2=\left(\fr 1 2 \la_8- \fr {f \om} {\sqrt 2 u v
}\right) (u^2+v^2). \ee The model contains two massive charged Higgs and two massless Higgs that are
identified to the Goldstone bosons of Y and W bosons. \be G_Y^-=\fr{\om \chi_2^- - v
\rho_3^-}{\sqrt{v^2+\om^2}}, \hs \hs  G_W^-=\fr{u \eta_2^-  -v \rho_1^-}{\sqrt{u^2+v^2}}. \ee
\end{itemize}

\subsection{\label{gsec}Gauge sector}

In this section, let us consider the gauge boson spectrum. The mass Lagrangian is given
as \bea \mathcal{L}_{gaugemass}&=&(0,0,\fr {\om}{\sqrt2})(g A_{a\mu}T_a-\fr 1 3 g_X  B_\mu -\fr 2 3
g_N  C_\mu)^2(0,0,\fr {\om}{\sqrt2})^T \crn &&+ (\fr {u}{\sqrt2},0,0)(g A_{a\mu}T_a-\fr 1 3 g_X
B_\mu +\fr 1 3 g_N  C_\mu)^2(\fr {u}{\sqrt2},0,0)^T \crn && + (0,\fr {v}{\sqrt2},0)(g
A_{a\mu}T_a+\fr 2 3 g_X  B_\mu +\fr 1 3 g_N  C_\mu)^2(0,\fr {v}{\sqrt2},0)^T \crn && +2 (g_N C_\mu
\La)^2 .\eea

Let us denote the following combinations \bea W_\mu^{\pm}=\fr{A_{1\mu} \mp i A_{2\mu}}{\sqrt 2},
\hs \hs Y_\mu^\mp=\fr{A_{6\mu} \mp i A_{7\mu}}{\sqrt 2}. \eea The non-Hermitian gauge bosons
$W_\mu^{\pm}, Y_\mu^\mp$ have the following masses \be M^2_W =\fr 1 4 g^2 (u^2 + v^2 ), \hs \hs
M^2_Y =\fr 1 4 g^2 (v^2 + \om^2). \ee It is worth noting that $A_{4\mu}$ and $A_{5\mu}$ gain the
same mass. Therefore, these vectors can be combined the following physical states \be X^0_\mu=
\fr{A_{4\mu}-i A_{5\mu}}{\sqrt 2}, \ee and its mass is given: \be M^2_X=\fr 1 4 g^2 (u^2 + \om^2).
\ee

There is a mixing among $A_{3\mu},A_{8\mu},B_\mu ,  C_\mu$ components. In the
basis of these elements, the mass matrix denoted by $M^2$ is given as
follows

\be
\fr{g^2}2\left(
  \begin{array}{cccc}
    \fr 1 2 (u^2 + v^2) &
     \fr{u^2 - v^2}{2\sqrt 3} &
    -\fr{t_1(u^2 +2 v^2)}{3} &
    \fr{t_2 (u^2 - v^2)}{3} \\
    \fr{u^2 - v^2}{2\sqrt 3} &
    \fr 1 6 (u^2 + v^2 + 4 \om^2) &
    -\fr {t_1(u^2 -2(  v^2  +\om^2))}{3\sqrt 3} &
    \fr {t_2(u^2 +v^2 +4\om^2)}{3\sqrt 3} \\
    -\fr{t_1(u^2 +2 v^2)}{3} &
    -\fr {t_1(u^2 -2(  v^2  +\om^2))}{3\sqrt 3} &
    \fr {2}{9}t_1^2 (u^2 +4 v^2  +\om^2) &
    -\fr {2}{9}t_1 t_2 (u^2 -2( v^2 +\om^2 )) \\
    \fr{t_2 (u^2 - v^2)}{3} &
    \fr {t_2(u^2 +v^2 +4\om^2)}{3\sqrt 3} &
    -\fr {2}{9}t_1 t_2 (u^2 -2( v^2 +\om^2 )) &
    \fr {2}{9}t_2^2 (u^2 + v^2 +4 ( \om^2+9 \La))  \\
  \end{array}
\right),\label{gm4}
\ee
where  $t_1\equiv g_X/g$, $t_2\equiv g_N/g$.

The mass matrix in (\ref{gm4}) contains one exact zero eigenvalue with the corresponding eigenstate
 as follows
\be
 A_\mu= \fr {\sqrt3} {\sqrt{3 + 4 t_1^2}}
\left(t_1 A_{3\mu} -  \fr{t_1}{\sqrt 3}A_{8\mu} +B_\mu\right).\label{basis1}
 \ee
It is worth to notice that $A_\mu$ is the combination of $A_{3\mu}, A_{8\mu}$, and $B_\mu$ without
contribution of the new gauge boson $C_\mu$. The factor $t_1$ can be expressed in term of the sine
of the weak mixing angle $s_W$ by identifying the coefficient of the $\overline{e}e\gamma$ vertex
with the electromagnetic coupling constant $e$, similarly as the analysis in \cite{tw}. We get \be
t_1=\fr{\sqrt{3} s_W}{\sqrt{3-4s_W^2}}. \ee The diagonalization of the mass matrix is done via
three steps. In the first step, in the base of $A_\mu,Z_\mu,Z^\prime_\mu, C_\mu $, the two
remaining $Z_\mu,Z^\prime_\mu$ gauge vectors are given by \bea && Z_\mu=\fr {\sqrt{3 +  t_1^2}}{
\sqrt{3 + 4 t_1^2}}A_{3\mu}+ \fr{ t_1 (\sqrt3 t_1 A_{8\mu} -3 B_\mu )} {\sqrt{3 +  t_1^2}\sqrt{3 +
4 t_1^2}}, \crn && Z^\prime_\mu=\fr{\sqrt 3}{\sqrt{3 +  t_1^2}} A_{8\mu} +\fr{t_1}{\sqrt{3 +
t_1^2}} B_\mu.\label{basis2} \eea

In this basis, the mass matrix $M^2$ becomes
\bea
\left(
  \begin{array}{cc}
    0 & 0 \\
    0 & M^{\prime 2} \\
  \end{array}
\right),
\eea
where $M^{\prime 2}$ is the $3 \times 3$ mixing mass matrix
of $Z_\mu,Z^\prime_\mu, C_\mu$ gauge bosons given as
\bea
\fr {g^2}2\left(
  \begin{array}{ccc}
    \fr {(3 + 4 t_1^2) (u^2 + v^2)}{ 2 (3 + t_1^2)}
    &  -\fr{\sqrt{3 + 4 t_1^2} ((-3 + 2 t_1^2) u^2 + (3 + 4 t_1^2) v^2)}
    {6 (3 + t_1^2)}
    & \fr {\sqrt{3 + 4 t_1^2} t_2 (u^2-v^2)}{3 \sqrt{3 + t_1^2}} \\
    -\fr{\sqrt{3 + 4 t_1^2} ((-3 + 2 t_1^2) u^2 + (3 + 4 t_1^2) v^2)}
    {6 (3 + t_1^2)}
    & \fr {(3 -2 t_1^2)^2 u^2 + (3 + 4 t_1^2)^2 v^2 + 4(3 + t_1^2)^2 \om^2}{18 (3 + t_1^2)}
    & \fr {t_2 ((3-2 t_1^2) u^2 + (3 + 4 t_1^2) v^2 + 4 (3 + t_1^2) \om^2)}{9\sqrt{3 + t_1^2}} \\
    \fr {\sqrt{3 + 4 t_1^2} t_2 (u^2-v^2)}{3 \sqrt{3 + t_1^2}}
    & \fr {t_2 ((3-2 t_1^2) u^2 + (3 + 4 t_1^2) v^2 + 4 (3 + t_1^2) \om^2)}{9\sqrt{3 + t_1^2}}
    & \fr 2 9 t_2^2 (u^2 + v^2 + 4 (\om^2 +9 \La^2)) \\
  \end{array}
\right). \label{gm3} \nonumber \\ \eea

The matrix given in (\ref{gm3}) can be diagonalized by using block
diagonalizing  method. In new basis $(\mathcal{Z}_\mu,\mathcal{Z}_\mu^\prime, Z^N_\mu)$,
the mass mixing matrix is given as
\bea
\left(
  \begin{array}{cc}
    A & 0 \\
    0 & m_{Z^N}^2 \\
  \end{array}
\right),
\eea
where $A$ is the $2\times 2$ matrix
\bea
A=\fr{g^2}2\left(
  \begin{array}{cc}
    \fr {(3 + 4 t_1^2) (u^2 + v^2)}{2 (3 + t1^2)}+\mathcal{O}(\fr{v^4}{\La^2})
    & -\fr {\sqrt{3 + 4 t_1^2} ((-3 + 2 t_1^2) u^2 + (3 + 4 t_1^2) v^2)}{6(3 + t_1^2)}
    +\mathcal{O}(\fr{v^2\om^2}{\La^2})\\
     -\fr {\sqrt{3 + 4 t_1^2} ((-3 + 2 t_1^2) u^2 + (3 + 4 t_1^2) v^2)}{6(3 + t_1^2)}
    +\mathcal{O}(\fr{v^2\om^2}{\La^2})
    & \fr {(3 - 2 t_1^2)^2 u^2 + (3 + 4 t_1^2)^2 v^2 + 4 (3 + t_1^2)^2 \om^2}{18(3 + t_1^2)}
    +\mathcal{O}(\fr{\om^4}{\La^2})\\
  \end{array}
\right),\crn
\eea
\be
m_{Z^N}^2 \simeq 4 g^2 t_2^2 \La^2 .
\ee

The new basis $(\mathcal{Z}_\mu,\mathcal{Z}_\mu^\prime, Z^N_\mu)$
is related to the basis
$(Z_\mu,Z^\prime_\mu,C_\mu)$ as following
\bea
\left(
  \begin{array}{c}
    \mathcal{Z}_\mu \\
    \mathcal{Z}_\mu^\prime \\
    Z^N_\mu \\
  \end{array}
\right)
=\left(
   \begin{array}{ccc}
     1 & 0 & -\epsilon_1 \\
     0 & 1 & -\epsilon_2 \\
     \epsilon_1 & \epsilon_2 & 1 \\
   \end{array}
 \right)
 \left(
  \begin{array}{c}
    Z_\mu \\
    Z^\prime_\mu \\
    C_\mu \\
  \end{array}
\right),\label{basis3} \eea where \bea \epsilon_1&=&-\fr{3 \sqrt{3 + 4 t_1^2} (u^2- v^2)} {2
\sqrt{3 + t_1^2} t_2 (u^2 + v^2 + 4 (\om^2 + 9 \La^2))},\crn \epsilon_2&=&\fr{(3 - 2 t_1^2) u^2 +
(3 + 4 t_1^2) v^2 + 4 (3 + t_1^2) \om^2} {2 \sqrt{3 + t_1^2} t_2 (u^2 + v^2 + 4 (\om^2 + 9
\La^2))}. \eea In the limit $\La \gg \om\gg u,v$, \be \mathcal{Z}_\mu \sim Z_\mu,\hs
\mathcal{Z}_\mu^\prime\sim Z^\prime_\mu,\hs
 Z^N_\mu \sim C_\mu.\ee
The new heavy gauge boson $Z^N_\mu$ is imbedded to the gauge group
 $U(1)_N$. It approximately does not mix to other gauge bosons.
$\mathcal{Z}_\mu$ and $\mathcal{Z}_\mu^\prime$ are mixing of the two physical field $Z^1_\mu$, $Z_{\mu}^2$.

\bea
Z^1_\mu &=& \cos \xi \mathcal{Z}_\mu - \sin \xi \mathcal{Z}_\mu^{\prime},\hs
Z^2_\mu = \sin \xi \mathcal{Z}_\mu + \cos \xi \mathcal{Z}_\mu^{\prime},\crn
m_{Z^1}^2 & \simeq &
\fr{g^2} {8}\left(u^2 + \om^2 + \fr{u^2 + 4 v^2 + \om^2}{3 - 4 s_W^2} \right.\crn
&-& \left. 4 \fr{\sqrt{c_W^4 u^4 + v^4 -c_{2W} v^2 \om^2 + c_W^4 \om^4 +
 u^2 (-c_{2W} v^2 + (-1 + 2 s_W^4) \om^2)}}{(3 - 4 s_W^2)}\right),\crn
m_{Z^2}^2& \simeq&
\fr{g^2} {8}\left(u^2 + \om^2 + \fr{u^2 + 4 v^2 + \om^2}{3 - 4 s_W^2} \right.\crn
&+& \left. 4 \fr{\sqrt{c_W^4 u^4 + v^4 -c_{2W} v^2 \om^2 + c_W^4 \om^4 +
 u^2 (-c_{2W} v^2 + (-1 + 2 s_W^4) \om^2)}}{(3 - 4 s_W^2)}\right),\label{mixZ2Zn}
\eea
where $\tan 2\xi=\fr{\sqrt{3 - 4 s_W^2} (c_{2W} u^2 - v^2)}
{((-1 + 2 s_W^4) u^2 -c_{2W} v^2 + 2 c_W^4 \om^2)}$.

If we assume $\om \gg u,v$, then $\tan 2\xi \rightarrow 0$. We get
\bea
Z^1_\mu &\sim &  \mathcal{Z}_\mu, \hs
m_{Z^1}^2\simeq \fr{g^2 (u^2 + v^2)} {4 c_W^2}, \crn
Z^2_\mu &\sim &  \mathcal{Z}_\mu^\prime,\hs
m_{Z^2}^2\simeq \fr{g^2c_W^2 \om^2}{(3 - 4 s_W^2)}.
\eea
The gauge boson $Z^1_\mu $ is identified as $Z_\mu $ in the standard model.

\section{\label{inf} Generation of Inflation in the 3-3-1-1 model}

 We would like to note that the scalar singlet $\phi$ is completely breaking $U(1)_N$. The vacuum
expectation value (VEV)  $<\phi>$ can stay at the same  scale as $\om$'s scale and the
interesting phenomenology of the model at TeV scale was studied in \cite{c5}.
In a different situation, this VEV can be very
high that can be integrated out from the low energy effective potential and a new gauge boson $Z_N$
decoupling from the gauge boson spectrum. In this part, we expect that the VEV of $\phi$ is very
high and  consider the singlet scalar $\phi$ plays the role of inflaton field.  The potential for
$\phi$ at the tree level can be read off from Eq. (\ref{scalar}) as
 \be V_\phi=\mu^2 \phi^\dagger
\phi + \la (\phi^\dagger \phi)^2 +\la_{10} (\phi^\dagger
\phi)(\rho^\dagger\rho)+\la_{11}(\phi^\dagger \phi)(\chi^\dagger \chi) +\la_{12} (\phi^\dagger
\phi)(\eta^\dagger \eta).\ee Due to the larger VEV of $\phi$, the interaction terms of the singlet
scalar Higgs and the ordinary 3-3-1 model Higgs triplets can be ignored. During inflation, we get
\be
 V_\phi=\mu^2 \phi^\dagger
\phi + \la (\phi^\dagger \phi)^2. \ee
This potential is taken part in the chaotic inflation.
However, the inflaton field has coupling to the matter fields which allow it to make the
transition to hot bing bang cosmology at the end of inflation, namely  \be \mathcal{L}\supset
4g_N^2 C^\mu C_{ \mu} \phi^2 +h'^\nu_{ab}\bar{\nu}^c_{aR}\nu_{bR}\phi.\label{mass1}\ee

We take into account quantum corrections to $V_\phi$ following the analysis of Coleman and Weinberg
\cite{weiberg} \be V_{\mathrm{eff}}=\fr 1 {64\pi^2}\sum_i [(-1)^{2J} (2J+1) m_i^4 \ln \fr
{m_i^2}{\Delta^2}],\ee where $i=\nu_{aR}, \phi, C_\mu, \chi,\rho,\eta$. \bea &&m_{\nu_{aR}}=-2
h'^\nu_{ab}\Phi; \hs m^2_\phi=2 (\mu^2+3 \la \Phi^2);\hs m^2_{C_\mu}=8 g_N^2 \Phi^2; \crn &&  \hs
m^2_\rho=2  \la_{10} \Phi^2; \hs m^2_\chi=2  \la_{11} \Phi^2; \hs m^2_\eta=2  \la_{12} \Phi^2.
 \eea
 We get
\bea V_{\mathrm{eff}}=&&\fr 1 {64\pi^2}\{[-32 \sum_i (h'^\nu_{ii})^4+192g_N^4+
4(\la_{10}^2+\la_{11}^2+\la_{12}^2) ]\Phi^4 \ln \fr {\Phi^2}{\Delta^2}\crn && +4(\mu^2+3 \la
\Phi^2)^2 \ln \fr {\mu^2+3 \la \Phi^2}{\Delta^2}\} \crn &&= \fr 1 {64\pi^2}\{a\Phi^4 \ln \fr {\Phi}{\Delta}
+4(\mu^2+3 \la \Phi^2)^2 \ln\fr {\mu^2+3 \la \Phi^2}{\Delta^2}\},  \eea
where
\be a=2[-32 \sum_i
(h'^\nu_{ii})^4+192g_N^4+ 4(\la_{10}^2+\la_{11}^2+\la_{12}^2) ].\label{a}\ee
  We identify the inflaton with
the real part of the $B-L$ Higgs field, $\Phi=\sqrt2 \mathcal{R}[\phi] $. In the leading- log
approximation, we obtain
\be V(\Phi)= V_{\mathrm{tree}}+V_{\mathrm{eff}}\simeq \fr{\mu^2}{2}\Phi^2+ \fr {\la}{4}\Phi^4+V_{\mathrm{eff}}.
\label{Vexact}\ee

We would like to remain that the inflation occurs as 
the inflaton slowly rolls to the minimal potential. The inflationary slow roll parameters are given
\cite{Linde} by \bea \epsilon(\Phi)=\frac{1}{2}m_P^2 \left( \frac{V^\prime}{V}\right)^2, \hs
\eta(\Phi)=m_P^2 \left( \frac{V^{\prime \prime}}{V}\right), \hs \zeta^2 (\Phi)=m_P^4 \frac{V^\prime
V^{\prime \prime \prime}}{V^2}, \eea where $m_P=2.4 \times 10^{18}$ GeV and a prime is denoted as a
derivative of $\Phi$. The slow roll condition means that $\epsilon (\Phi) \ll 1, \mid\eta(\Phi)\mid
\ll 1, \zeta(\Phi) \ll 1$. In this limit, the spectral index $n_s$, the tensor to scalar ratio $r$
(a canonical measure of gravity wave from inflation) and the running index $\al $ can be written as
\bea n_s=1-6\epsilon+2\eta, \hs r=16\epsilon, \hs \al =16\epsilon\eta-14\epsilon^2-2\zeta^2. \eea
The spectrum index $n_s$ is estimated by BICEP2 experiment \cite{bicep2}, Planck \cite{planck} and
WMAP9 \cite{wmap9} measurements. It is  closed to 0.96. The tensor to scalar ratio is proven by
BICEP2 \cite{bicep2}, $r = 0.20^{+0.07}_{-0.05}$ while the Planck and WMAP9 experiments gave the
bound $r < 0.11 (0.12).$

The  number of
e-folds is given by \bea N= \int^{\Phi_0}_{\Phi_e}\frac{V d \Phi}{V^\prime}, \eea where $\Phi_e$ is
the inflaton value at the end of inflation and defined by max$(\epsilon(\Phi), \eta(\Phi),\zeta
(\Phi))=1$. $\Phi_0$ is the inflation value at the horizon exit. The value of $N$ is around $50-60$
and depends on the energy scale during inflation.

The amplitude of the curvature perturbation is given as follows \bea \triangle^2_\mathcal{R}=
\frac{V}{24\pi^2m_P^4\epsilon(\Phi)}, \eea The value of curvature perturbation should satisfy the
Planck measurement \cite{plancka}:  $\triangle^2_\mathcal{R}=2.215 \times 10^{-9}$ at the scale
$k_0=0.05 Mpc^{-1}$.

\begin{figure}[!h]
\begin{center}
\includegraphics[width=15cm,height=8cm]{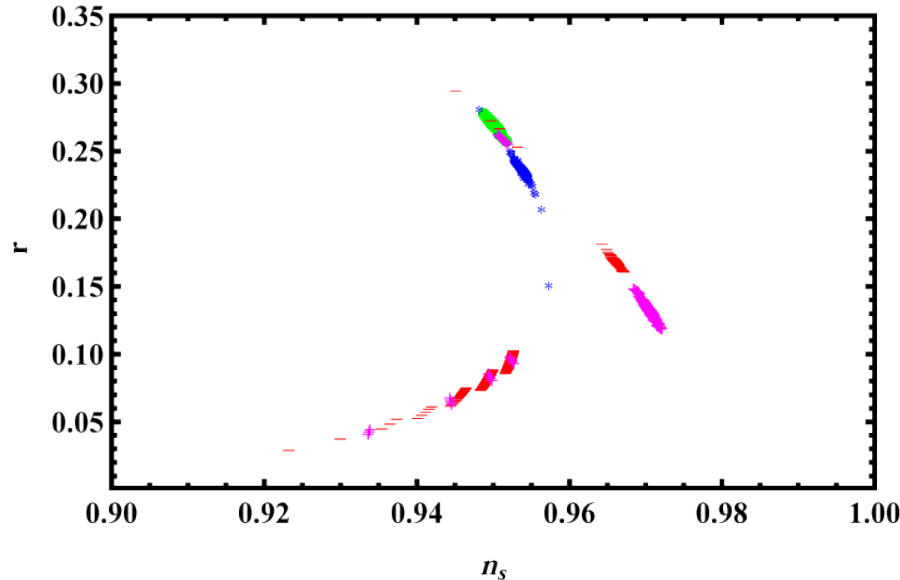}
\includegraphics[width=15cm,height=8cm]{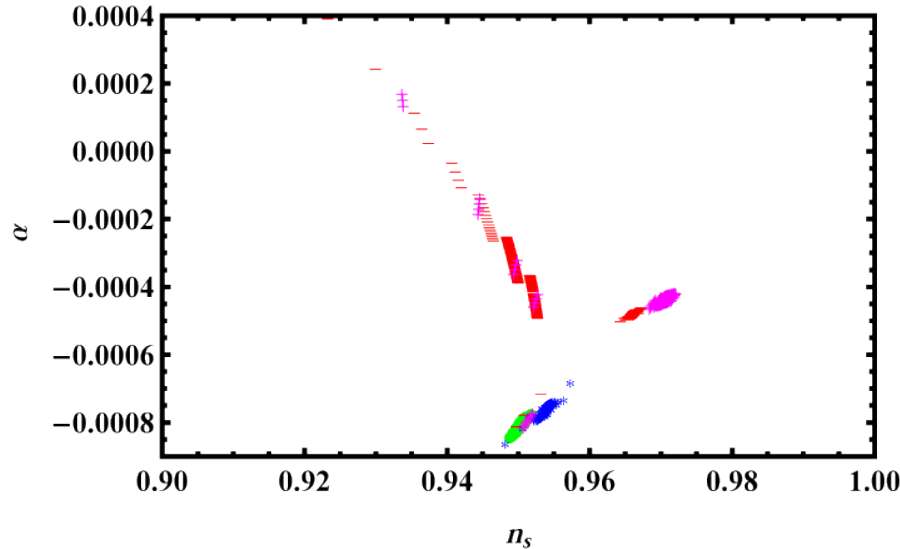}
\caption[]{\label{figI}{$r$  vs. $n_s$ (upper panel) and $\alpha$ vs.  $n_s$(lower panel)
for $\Delta= 0.1 m_P$ (green dot), $\Delta= 30 m_P$ (red minus),
$\Delta= 50 m_P$ (pink plus), and $\Delta= 500 m_P$ (blue asterisk) in the range of  $-10^3<a'<10^3$,
with the number of e-folds $N=60$.}}
\end{center}
\end{figure}

Let us study parameter space of $\mu,\la, \Delta, a$ appeared in the potential $V(\Phi)$.
If $\mu^2  \gg \la \Phi^2$ or $\mu^2  \sim \la \Phi^2$, both $\triangle^2_\mathcal{R}$ and $r$
either one of them is not in agreement with the Planck and WMAP9 experimental results.
For example, taking $\Delta=30 m_P$, and random values of other
parameters $10^{-10} m_P^2<|\mu^2|<10^4 m_P^2$, $10^{-15}<|a|<10^3$, $10^{-10}<|\la|<1$ we get
$|\triangle^2_\mathcal{R}|>10^3$.
If we assume that $\mu^2 \ll \la \Phi^2$, the potential (\ref{Vexact}) can be rewritten
in simple form
\be V(\Phi)= \la' (\Phi^4+ a'\Phi^4 \ln\fr{\Phi} {\Delta} ),
\ee
where \be\la'=\fr{\la}4, \hs a'= \fr{a+ 72 \la^2 }{16\pi^2\la}.\ee
The coupling constant $\la$ is determined to satisfy the constraint on
$\triangle^2_\mathcal{R}$, while as the predictions for $n_s, r, \alpha$
are given for fixed values of $a', \Delta$. Fig. \ref{figI} shows the predicted values
of $n_s$, $r$  and $\alpha$ for $\Delta= 0.1 m_P$ (green), $\Delta= 30 m_P$ (red),
$\Delta= 50 m_P$ (pink), and $\Delta= 500 m_P$ (blue) in the range of  $-10^3<a'<10^3$
with the number of e-folds $N=60$. We can see that for $\Delta= 0.1 m_P$ and $\Delta= 500 m_P$,
$r$ runs out of experimental region for almost values of $a'$ in the range $-10^3<a'<10^3$.
For $\Delta= 30 m_P$, we need to require $a'<-36$ or $a'>6$ to make sure $n_s$ and $r$ are in agreement
with experimental results \cite{Planck13}, $n_s \in (0.94,0.98)$, and $r\in (0.001,0.15)$. 

If we vary $a', \Delta$ in the parameter region satisfying experimental results,
the order of $<\Phi>$ and the inflaton mass mostly does not change.
From now on we take $a'=-10^2, \Delta= 30 m_P$ for the below numerical calculations.

From the minimal potential condition, we get $<\Phi> \simeq 23.6 m_P$.
The inflaton mass is calculated by the second derivative of the effective
potential at the minimum.  For $Z^N$ and $\nu_{kM}$, the mass arises from
(\ref{mass1}) with notice that $\Phi=\sqrt2 \mathcal{R}[\phi]$.
We obtain
\be m_\Phi=\sqrt{V''(\Phi)}\left|_{\Phi=<\Phi>} \right. \simeq 2.67 \times10^{13} \mathrm{GeV},\hs  m_{Z^N}=2 g_N <\Phi>,
\hs  m_{\nu_{iR}}=-\sqrt2 h'^\nu_{ii} <\Phi>.  \ee

Now let us calculate the reheating temperature.
In this model, the inflaton couples to pair of Higgs, pair of gauge boson
$Z^N$ and pair of Majorana neutrinos.
We assume that $m_\Phi<m_{Z^N}$, hence the inflaton cannot decay into pair of $Z^N$.
The inflaton can decay into pair of Higgs  with the decay rate
\bea \Gamma (\Phi\rightarrow hh)=\fr {\la^2_{10;11;12}<\Phi>^2}{32 \pi m_\Phi}.\eea
If the mass condition is allowed, the inflaton can decay into pair of $\nu_{iR}$ 
\bea \Gamma (\Phi\rightarrow \nu_{iR}\nu_{iR})=\fr {(h'^\nu_{ii})^2 m_\Phi}{16 \pi}.\eea
If $|\la_{10;11;12}| \gg \fr{\sqrt2 |h'^\nu_{ii}|m_\Phi }{<\Phi>} $, we get
$\Gamma (\Phi\rightarrow hh)\gg \Gamma (\Phi\rightarrow \nu_{iR}\nu_{iR})$.
The inflaton dominantly decays into pair of Higgs, therefore the reheating temperature is
estimated as
 \be T_R= \left(\fr{90}{\pi^2 g^*}\right)^{\fr 1 4} (\Ga_\Phi m_P)^{\fr 1 2}\sim
 10^{20} \mathrm{GeV} \times |\la_{10;11;12}|, \ee
  where $g^* =106.75$ is the number of degrees of the freedom active at the temperature of the
 asymmetry production.
The constraint $\Gamma (\Phi\rightarrow hh)\ll m_\Phi$ requires $|\la_{10;11;12}| \ll
\fr{\sqrt{32 \pi}m_\Phi}{<\Phi>}\sim 10^{-6}$. We find the limit   $T_R (\mathrm{max})<10^{14}$ GeV.
Taking $|\la_{10;11;12}|\sim 10^{-11}$ then $T_R\sim 10^{9}$ GeV satisfying the
 upper bound on reheating temperature to prevent gravitinos problem. 
In this case the thermal leptogenesis scenario may work to explain the baryon asymmetry.

In other case, we assume
\be  |h'^\nu_{11}| \ll\fr {m_\Phi}{\sqrt2 <\Phi>} \sim 3.33\times 10^{-7}< |h'^\nu_{22}|\sim |h'^\nu_{33}|,
 \label{const-h11}\ee therefore,
\be m_{\nu_{1R}}\ll m_\Phi <  m_{\nu_{2R}}\sim m_{\nu_{3R}}.   \ee

If $|\la_{10;11;12}|\ll \fr{|h'^\nu_{11}|m_\Phi }{<\Phi>}$, we get $\Gamma (\Phi\rightarrow hh)\ll \Gamma (\Phi\rightarrow \nu_{1R}\nu_{1R})$.
When $\la_{10;11;12}$ are negligibly small, the inflaton dominantly decays into pair of
$\nu_{1R}$.
 The produced reheating temperature is given as
 \be T_R= \left(\fr{90}{\pi^2 g^*}\right)^{\fr 1 4} (\Ga_\Phi m_P)^{\fr 1 2}\sim 10^{14} \times |h'^\nu_{11}|.\label{TRnon}\ee
This temperature is much lower than the RH neutrino mass since $<\Phi>$ is at Planck value.
 We can apply  non-thermal leptogenesis scenario, in which the $\nu_{1R}$ is produced through the direct non-thermal decay of the inflaton $\Phi$.

\section{\label{lept}Leptogenesis  in the 3-3-1-1 model}

 First, we consider the scalar sector.
 The scalar mass spectrum  is considered in \cite{c4, c5},
in which $-f\sim \om \sim \La \gg u\sim v$.
 In this work we assume $\Lambda \gg \om\gg u\sim v$, the considered model contains
 \begin{itemize}
\item  There are $9$ Goldstone bosons $A_4, G_Z, G'_Z, G_X, G_X^*,  G_W^\pm,G_Y^\pm $, their interactions
 can be gauged away by a unitary transformation.
 \item One higgs gains mass at the electroweak breaking scale.
 This  is the lightest massive Higgs bosons $H$ and is identified as SM Higgs.
 \item There are 9 new Higgs bosons namely,
 $A, H_1, H_2,H_4^\pm, H_5^\pm , H^\prime,
 H^{\prime*} $, which are heavy  at the
 $\om$ scale, while the mass of $H_3$ is proportional to $\La$.
 \end{itemize}
 
 In the gauge sector, let us collect the new gauge bosons beyond the SM.
 In the limit $\La\gg \om \gg u,v$, we get
\begin{itemize}
 \item  One super heavy gauge boson
 $Z_\mu^N \sim C_\mu$ with the mass  $m_{Z^N}^2 \simeq 4 g_N^2\La^2$.
 \item All the other new
 gauge bosons, $Z_\mu^2,  X_\mu^0, X_\mu^{0*}, Y_\mu^\pm$,
 have  mass in order $\cal{O}(\om)$.
 \end{itemize}
 The lepton number of particles are considered in \cite{c4,c5}. In
 particularly, the SM  particles have a lepton number as usual.
 The  new particles $(G_X, H^{\prime *}, H_4^-, G_Y^-, X^0, Y^- )$
 have the lepton number equal to one, their  complex conjugate have the lepton number equal to minus one while
 the remaining Higgs and gauge bosons have zero lepton number.

 Now in order to account for leptogenesis, we have to verify the
 lepton number violating interactions. Seeing that the lepton
 number $L$ and baryon number $B$ are conserved by VEVs of $\eta, \chi, \rho$
 as mentioned in \cite{c4}. All interaction terms appeared after
 symmetry breaking in the considered model are conserved the
 lepton number.
 Hence it is clear that $B$, $L$ violating number interactions should
 be broken in other way in order to explain neutrino
 mass and mixing as well as
 the matter-antimatter asymmetry of the Universe.
 The lepton number only can be violated in the
 interactions of Majorana neutrinos with non-zero lepton number particles.

Let us remind the seesaw mechanism that explains the tiny neutrino
mass and large mixing. The Lagrangian relevant to the neutrino mass has
a form as
\bea L_{\nu-mass}=
(h^\nu_{ab}\bar{\psi}_{aL}\eta\nu_{bR}+
  h'^\nu_{ab}\bar{\nu}_{aR}^c  \nu_{bR}\phi
  + \mathrm{H. c})\label{neumass}.\eea
  The left handed neutrinos couple to the right handed neutrinos
  through the first term of the Eq. (\ref{neumass}) and have a Dirac
  mass as
  \bea [m_\nu^D ]_{ab}= -\frac{h^\nu_{ab}}{\sqrt{2}}u, \eea
  while the right handed neutrinos couples to themselves through the
  second term given in the Eq. (\ref{neumass}) and have
  a Majorana mass as
   \bea [m_\nu^M ]_{ab}=-\sqrt{2}h_{ab}^\prime \Lambda.\eea
Hence, we can explain the smallness of the light neutrino masses
 via a
type I seesaw mechanism \cite{c4} and predict six Majorana
neutrinos as mass eigenstates, three heavy neutrinos $\nu_{iM}$
and three light neutrinos $\nu_{iE}$,
 \be \nu_{iM}=\nu_{iR}+\nu_{iR}^c; \hs m_{\nu_{M}}
 = m_\nu^M=-\sqrt{2}h^\prime\Lambda, \label{Ma1}\ee
\be \nu_{iE} = \nu_{iL}+\nu_{iL}^c; \hs
m_{\nu_{E}}^\mathrm{eff}=-m_{\nu}^D (m_{\nu}^M)^{-1} (m_{\nu}^D)^T
= \fr {u^2}{2\sqrt2\La}h^\nu (h'^\nu)^{-1} (h^\nu)^T .\label{Ma2} \ee

We note that the considered model also contains three new neutral
fermions $N_{aR}$. They obtain the Majorana masses \cite{c4} via
an effective interaction as
 \be
 \fr{\la_{ab}}{M}(\bar{\psi}^c_{aL})_m (\psi_{bL})_n
 (\chi_m \chi_n)^* +\mathrm{H. c}.
 \ee
The Majorana masses of the neutral fermions $N_{aR}$ are given
\bea[ m_{N_R}]_{ab} = -\frac{\lambda_{ab}\om^2}{M} \eea
 and  the Majorana fermion states are
  \bea
 N_i&=&N_{iR}+N_{iR}^c.
\label{Ma3} \eea
 Based on the Majorana fermion states given in Eqs. (\ref{Ma1}), (\ref{Ma2})
 and (\ref{Ma3}), we can rewrite the Lagrangian $\mathcal{L}_{\nu_{R}}$
 including the Yukawa terms in Eq. (\ref{neumass})
 and the gauge-fermion interaction
$\bar{\nu}_{iR}i \gamma^\mu D_\mu \nu_{iR}$  as follows
 \bea \mathcal{L}_{\nu_{R}}&=&(h^\nu_{ab}\bar{\psi}_{aL}\eta\nu_{bM}+
  h'^\nu_{ab}\bar{\nu}_{aM} P_R \nu_{bM} \phi
  -\fr 1 2 [m_{\nu}^M]_{ab} \bar{\nu}_{aM} P_R \nu_{bM}
  + \mathrm{H. c})\crn
 &&+ g_N \bar{\nu}_{iM}\gamma^\mu P_R \nu_{iM} Z_\mu^N
 +\bar{\nu}_{iM}i\gamma^\mu  \partial_\mu P_R\nu_{iM}+H.c. \label{LnuM}
 \eea

 To rely on Higgs physical states mentioned above, we  obtain the
 physical interaction terms that violate the lepton number. In particularly
 the lepton violating
 interactions appeared in Eq. (\ref{LnuM}) are:  $\bar{e} P_R \nu_M H_5^-,
 \bar{N} P_R H^\prime \nu_M$.

We would like to emphasize that the lepton number violating terms
also appear via the interactions of the light Majorana neutrinos,
namely, $\bar{e}\nu_E W^-, \bar{N}\nu_E X^{0*}$. However these
interactions  do not generate baryon asymmetry by \cite{washout}.
In brief, this model contains the lepton number violating
 interactions, which are  $\bar{e} P_R \nu_M H_5^-,
 \bar{N} P_R H^\prime \nu_M$,  $\bar{e}\nu_E W^-, \bar{N}\nu_E X^{0*}$.
 We consider leptogenesis scenario at the temperature $T_{\Gamma}$ satisfying
 $T_{\Gamma}\gg 1 $TeV. It implies that
 only $\nu_M$ can generate lepton asymmetry.

 Before calculating the CP asymmetry of $\nu_{aM}$,
 for convenience, we list all non-zero couplings of fermions
 appearing in loop diagram of $\nu_M \rightarrow e+ H_5^+$ and
 $\nu_M \rightarrow N+ H'^*$.

 \begin{table}[h]
 \bc
 \begin{tabular}{|c|c|c|c|}
   \hline
   vertex & coupling & vertex & coupling \\ \hline
   $\bar{\nu}_{aE}\nu_{bM}H$ & $\fr{u h^\nu_{ab}}{\sqrt 2 \sqrt{u^2 + v^2}} P_R$
   & $\bar{\nu}_{aE}\nu_{bM}H_1$ & $-\fr{v h^\nu_{ab}}{\sqrt 2 \sqrt{u^2 + v^2}}
    P_R$  \\\hline
   $\bar{e}_a\nu_{bM}H_5^-$ & $\fr{v h^\nu_{ab}}{\sqrt{u^2 + v^2}} P_R$
   & $\bar{N}_a\nu_{bM}H'$ & $\fr{\om h^\nu_{ab}}{\sqrt{u^2 + \om^2}} P_R$
    \\\hline
   $\bar{\nu}_{aM}\nu_{bM}H_3$ & $\fr{h'^\nu_{ab}}{\sqrt 2} P_R
   +\fr{h'^{\nu *}_{ba}}{\sqrt 2} P_L $
   & $\bar{\nu}_{aM}\nu_{bM}Z^N_\mu$ & $g_N \ga^\mu P_R$ \\\hline
   $\bar{\nu}_{aE} e_b H_5^+$ & $\fr{u h^e_{ab}}{\sqrt{u^2 + v^2}} P_R$
   &$\bar{N}_a e_b H_4^+$ & $\fr{\om h^e_{ab}}{\sqrt{v^2 + \om^2}} P_R
   + \fr{v\om \la_{ab}}{\sqrt 2\sqrt{v^2 + \om^2} M} P_L$\\\hline
   $\bar{N}_{a}\nu_{bE} H'$ & $\fr{u\om\la_{ab}}{\sqrt 2\sqrt{u^2 + \om^2} M} P_L$
   &$\bar{e}_{a}\nu_{bE} W_\mu^-$  & $-\fr{g\ga^\mu}{\sqrt 2}P_L$ \\\hline
   $\bar{e}_a e_b H$ & $\fr{v h^e_{ab}}{\sqrt 2\sqrt{u^2 + v^2}} P_R
   +\fr{v h^{e*}_{ba}}{\sqrt 2\sqrt{u^2 + v^2}} P_L $
   & $\bar{e}_a e_b H_1$ & $\fr{u h^e_{ab}}{\sqrt 2\sqrt{u^2 + v^
   2}} P_R + \fr{u h^{e*}_{ba}}{\sqrt 2\sqrt{u^2 + v^2}} P_L$ \\\hline
   $\bar{e}_a e_b A_\mu$ & $g s_W \ga^\mu$ & $\bar{e}_a e_b Z^k_\mu
   $& $\ga^\mu (g_{kV} -g_{kA}\ga^5 ), k=1,2,N$ \\\hline
   $\bar{\nu}_{aE}N_{b}X_\mu$ & $-\fr{g\ga^\mu}{\sqrt 2}P_L$
   & $\bar{e}_{a}N_{b}Y_\mu^-$ & $-\fr{g\ga^\mu}{\sqrt 2}P_L$\\\hline
   $\bar{N}_a N_b Z^2_\mu $& $ \fr {g c_W}{\sqrt{3-4 s_W^2}} \ga^\mu P_L $
   & $\bar{N}_a N_b Z^N_\mu $& $ \fr {2}{3}g_N \ga^\mu P_L $\\\hline
   $\bar{N}_a N_b H_2 $& $ \fr {\om \la_{ab}}{2M}  P_L+ \fr {\om \la_{ba}^*}{2M}  P_R$
   &  &  \\
   \hline
 \end{tabular}
 \caption{Non-zero couplings of fermions appearing in loop diagram of
  $\nu_M \rightarrow e+ H_5^+$ and
 $\nu_M \rightarrow N+ H'^*$ .} \label{coupling1}
 \ec
 \end{table}

\begin{figure}[!h]
\begin{center}
\includegraphics[width=14cm,height=12cm]{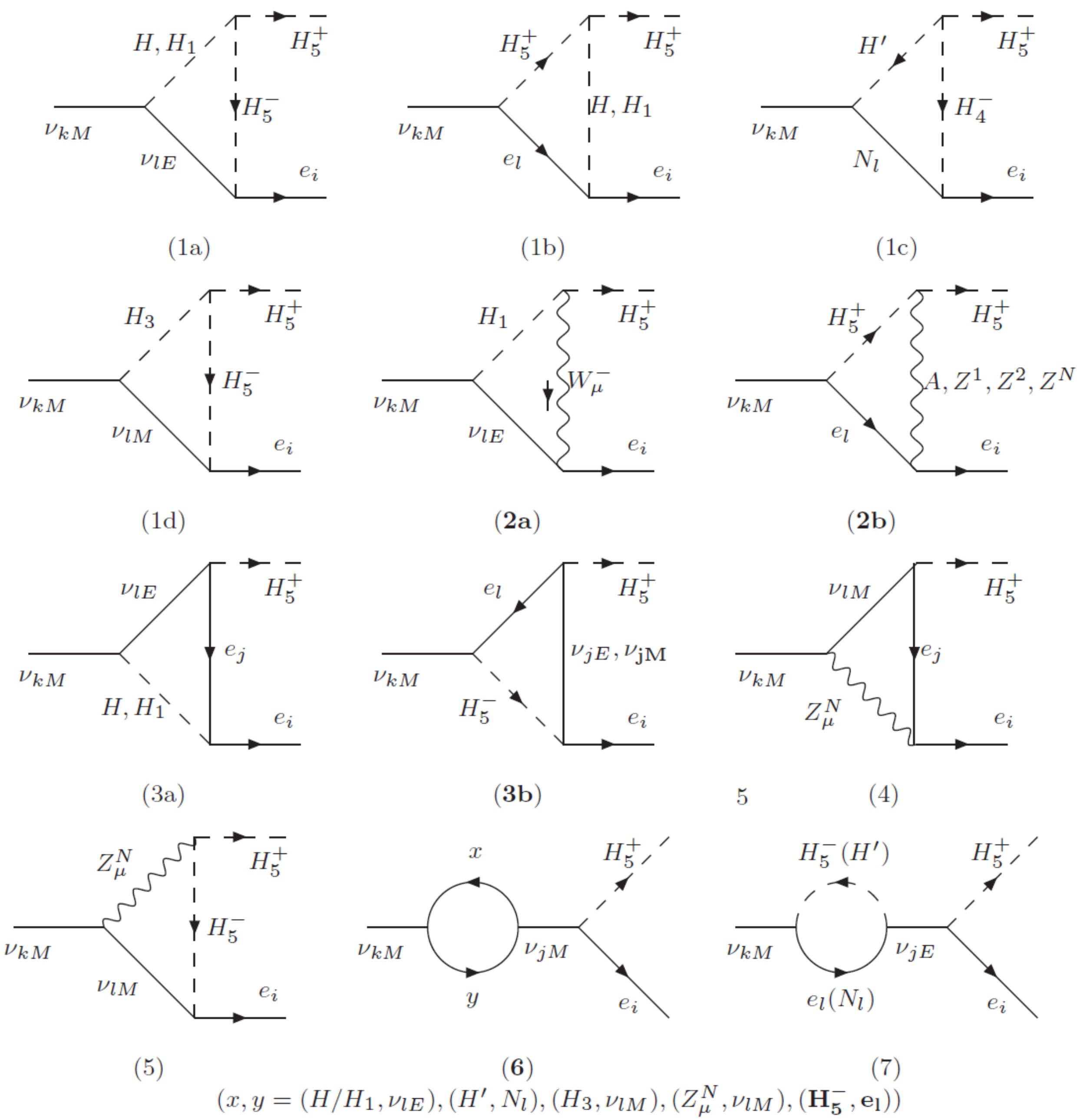}
\caption[]{\label{fig1} One-loop diagram contributing to the asymmetry from the
decay  $\nu_M \rightarrow e+ H_5^+$.}
\end{center}
\end{figure}

 All possible one-loop diagrams, which can contribute to the
 CP asymmetry from the decay
 $\nu_M \rightarrow e+ H_5^+$ are listed in Fig. \ref{fig1}.
 The interference of the tree level and one-loop level
 (2a, 2b), (3b) with the propagator $\nu_{jM}$, (6) with the propagator ($H_5^-, e_l$)
 gives dominated contribution to the CP asymmetry. We obtain

 \bea
 \varepsilon^{i(1)}_{\nu_{kM}} &=& \fr {\Gamma(\nu_{kM}\rightarrow e_i+H_5^+)
 -\Gamma(\nu_{kM}\rightarrow \overline{e}_i+H_5^-)}{2\Gamma_{\nu_{kM}} }\crn
  &\simeq& \fr {1}{8\pi C_0} \left[\fr 1 2 g \la_{W^+H_1H_5^-}
 + e\la_{A H_5^+H_5^-}+(g_{1V}+g_{1A})
 \la_{Z_1 H_5^+H_5^-}+(g_{2V}+g_{2A})\la_{Z_2 H_5^+H_5^-}\right.\crn
 &&+ \left. (g_{NV}+g_{NA})\la_{Z_N H_5^+H_5^-}(4-2 g_z log [1+1/g_z])\right]
 s_\beta^2 \sum_{ l}\mathrm{Im}[ h^{\nu*} _{ik} h^\nu _{lk}]
\crn
 &&+ \fr{s_\beta^4}{8\pi C_0} \sum_{j} \sqrt{g_j}\left[1-(1+g_j)log[1+1/g_j]+ (1-g_j)^{-1}\right]
 \mathrm{Im}[( h^{\nu\dag}h^\nu)_{kj}h^{\nu*} _{ik} h^\nu _{ij}],\label{epsilon1}
 \eea
 where $\Gamma_{\nu_{kM}}$ is the total decay rate of $\nu_{kM}$ at tree level,
 \bea
 &&g_z=\fr {m_{Z_N}^2}{m_{\nu_{kM}}^2}, \hs g_j= \fr {m_{\nu_{{jM}}}^2}
 {m_{\nu_{kM}}^2},\hs C_0=(2+s_\beta^2)\sum_i |h^\nu_{ik}|^2=(2+s_\beta^2)( h^{\nu\dag}h^\nu)_{kk} , \hs t_\beta=v/u,\crn
 &&  \la_{W^+H_1H_5^-}=\fr g 2 ,\hs \la_{A H_5^+H_5^-}= -e, \hs
 \la_{Z_1 H_5^+H_5^-}=-\fr{g c_{2W}}{2 c_W}, \hs \la_{Z_2 H_5^+H_5^-}=-\fr{g (c_\beta^2- c_{2W}s_\beta^2)}{2 c_W \sqrt{3-4 s_W^2}}, \crn
 &&\la_{Z_N H_5^+H_5^-}= -\fr{g_N c_{2\beta} }3, \hs
 (g_{1V}+g_{1A})=-(g_{2V}+g_{2A})=\fr {g c_{2W}}{2 c_W} ,  \hs (g_{NV}+g_{NA})= \fr{2g_N}3.
 \eea
Here we ignore the mixing between $Z_\mu^1$ and $Z_\mu^2$ since $\om>> u,v$.

\begin{figure}[!h]
\begin{center}
\includegraphics[width=14cm,height=12cm]{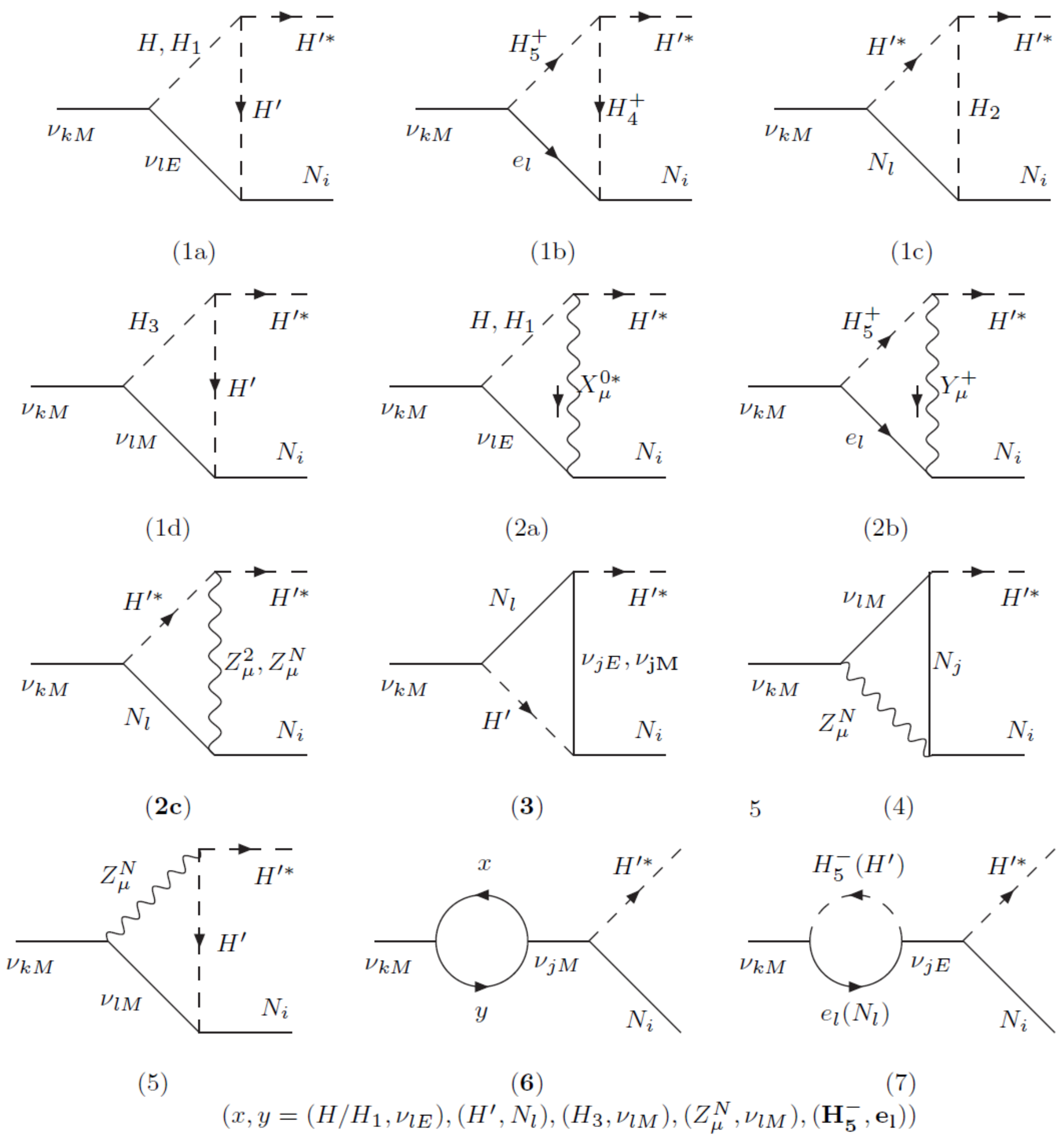}
\caption[]{\label{fig2} One-loop diagram contributing to the asymmetry from the
 decay  $\nu_M \rightarrow N+ H'^*$.}
\end{center}
\end{figure}

Now we consider CP asymmetry of the decay $\nu_M \rightarrow N+ H^{'*}$.
All possible loop diagrams are listed in the Fig. \ref{fig2}.
The interference of the tree level and one-loop level
 (2c), (3) with the propagator $\nu_{jM}$, (6) with the propagator ($H_5^-, e_l$)
 gives dominated contribution to the CP asymmetry. We obtain
 \bea
 \varepsilon^{i(2)}_{\nu_{kM}} &=& \fr {\Gamma(\nu_{kM}\rightarrow N_i+H'^*)
 -\Gamma(\nu_{kM}\rightarrow N_i+H')}{2\Gamma_{\nu_{kM}} }\crn
&\simeq& \fr {1}{8\pi C_0} \left[\fr{g c_W}{\sqrt{3-4s_W^2}}\la_{Z_2 H'^*H' }+ \fr 2 3 g_N\la_{Z_N H'^*H' }(4-2 g_z log [1+1/g_z])\right] \sum_{l}
\mathrm{Im}[ h^{\nu*} _{ik} h^\nu _{lk}]
\crn
 &&+\fr{1}{8\pi C_0} \sum_{j} \sqrt{g_j}\left[1-(1+g_j)log[1+1/g_j]+ s_\beta^2(1-g_j)^{-1}\right]
 \mathrm{Im}[( h^{\nu\dag}h^\nu)_{kj}h^{\nu*} _{ik} h^\nu _{ij}],\label{epsilon2}
 \eea
 where \be \la_{Z_2 H'^*H' }=-\fr{g c_W}{\sqrt{3-4s_W^2}}, \hs \la_{Z_N H'^*H' }=\fr {g_N}3.
 \ee
We would like to notice that since the coupling $\la_{\overline{e}_a \nu_{bM}H_5^-}=s_\beta h^\nu_{ab} P_R$ while $\la_{\overline{N}_a \nu_{bM}H'}\simeq h^\nu_{ab} P_R$, the factors $s_\beta^2,s_\beta^4 $ appear in (\ref{epsilon1})
while $s_\beta^0,s_\beta^2 $ appear in (\ref{epsilon2}).
In this work we take $u\sim v$ and thus $s_\beta=1/\sqrt2$.

 Let us comment on neutrino mass and mixing. The light neutrino mass matrix is given Eq. (\ref{Ma2}).  In order
 to diagonal this matrix, we have to use the $U$ matrix. It is nice to note that the lepton mixing matrix
 was studied by the Pontecorvo-Maki-Nakagawa-Sakata (PMNS). The standard form of this mixing matrix is
 given
 \bea
U_{\rm PMNS}=\left(%
\begin{array}{ccc}
  c_{12}c_{13} & s_{12}c_{13} & s_{13}e^{-i\delta} \\
  -s_{12}c_{23}-c_{12}s_{23}s_{13}e^{i\delta} & c_{12}c_{23}-s_{12}s_{23}s_{13}e^{i\delta} & s_{23}c_{13} \\
  s_{12}s_{23}-c_{12}c_{23}s_{13}e^{i\delta} & -c_{12}s_{23}-s_{12}c_{23}s_{13}e^{i\delta} & c_{23}c_{13} \\
\end{array}%
\right).
 \eea
where $c_{ij}= \cos \theta_{ij}, s_{ij}= \sin \theta_{ij}$ and the values of $\theta_{ij}$ are
determined by the global analysis \cite{neutrino}, namely \bea \sin^2 \theta_{23}=0.466^{+0.073,
0178}_{-0.058, 0.135}; \hs  \sin^2 \theta_{12}= 0.312^{+0.019, 0.063}_{-0.0018, 0,049}; \hs  \sin^2
\theta_{13} = 0.016\pm 0.010 (\leq 0.046). \nonumber \\ \eea
 $\delta$ is unknown CP violating Dirac phase.

 On the other hand,
the square of charged lepton mass matrix and light
neutrino mass matrix
 are diagonalized by two unitary transformations
 \bea
U_l^\dag M_l^+ M_l U_l = Diag(m_e^2, m_\mu^2, m_\tau^2); \hs U^T_\nu m_{\nu E}^{eff}U_\nu = Diag
(m_{\nu_1}, m_{\nu_2}, m_{\nu_3})
 \eea
 The $U_{\rm PMNS}$ is defined as
 \bea
U_{\rm PMNS} P =U^\dag_l U_\nu,
 \eea
 where
\bea
P= \left(%
\begin{array}{ccc}
  1 & 0 & 0 \\
  0 & e^{i \sigma} & 0 \\
  0 & 0 & e^{i \rho} \\
\end{array}%
\right), \eea
where  $\rho, \sigma$ are CP violating Majorana phases.
If we ignore the mixing between the charged lepton, then we can get \bea U_\nu =
U_{\rm PMNS} P. \eea
We assume that $h^{\prime \nu}= Diag (h_{11}^{\prime \nu},h_{22}^{\prime \nu},h_{33}^{\prime \nu}
)$ then $m_{\nu}^M=Diag(m_{\nu_{1M}},m_{\nu_{2M}},m_{\nu_{3M}})$.
Using the analysis in \cite{R-ortho}, the most general  $h^\nu$ matrix
is given by
\be h^\nu=\fr{\sqrt2}{u} Diag(\sqrt{m_{\nu_{1M}}},\sqrt{m_{\nu_{2M}}},\sqrt{m_{\nu_{3M}}}). R.
Diag(\sqrt{m_{\nu_1}},\sqrt{m_{\nu_2}},\sqrt{m_{\nu_3}}). U^\dag, \label{hnu} \ee
where $R$ is orthogonal matrix expressed in terms of arbitrary
complex angles $\widehat{\theta}_1,\widehat{\theta}_2,\widehat{\theta}_3$ as following
\be R=\left(
        \begin{array}{ccc}
          \widehat{c}_2\widehat{c}_3 & -\widehat{c}_1\widehat{s}_3-\widehat{s}_1\widehat{s}_2\widehat{c}_3 &
          \widehat{s}_1\widehat{s}_3-\widehat{c}_1\widehat{s}_2\widehat{c}_3 \\
          \widehat{c}_2\widehat{s}_3 &
          \widehat{c}_1\widehat{c}_3-\widehat{s}_1\widehat{s}_2\widehat{s}_3 &
          -\widehat{s}_1\widehat{c}_3-\widehat{c}_1\widehat{s}_2\widehat{s}_3 \\
          \widehat{s}_2 & \widehat{s}_1\widehat{c}_2 & \widehat{c}_1\widehat{c}_2 \\
        \end{array}
      \right),
\ee
where $\widehat{c}_i = \cos \widehat{\theta_i}, \widehat{s}_i = \sin\widehat{\theta_i}$,  $i=1,2,3.$

From the Eq. (\ref{hnu})  $h^{\nu\dag} h^\nu$ has the form
\be h^{\nu\dag} h^\nu=\fr{ 2}{u^2} U. Diag(\sqrt{m_{\nu_1}},\sqrt{m_{\nu_2}},\sqrt{m_{\nu_3}}).R^\dag.
Diag(m_{\nu_{1M}},m_{\nu_{2M}},m_{\nu_{3M}})
.R.Diag(\sqrt{m_{\nu_1}},\sqrt{m_{\nu_2}},\sqrt{m_{\nu_3}}). U^\dag. \label{hnu1} \ee

For the light neutrinos masses, we fit the experimental results
\be \Delta m_{\nu_{12}}^2=m_{\nu_2}^2-m_{\nu_1}^2=7.53 \times 10^{-5}\mathrm{eV^2}, \ \Delta m_{\nu_{23}}^2=m_{\nu_3}^2-m_{\nu_2}^2=2.44 \times 10^{-3}\mathrm{eV^2}. \ee
The  asymmetry $\varepsilon^{i(1)}_{\nu_{kM}}, \varepsilon^{i(2)}_{\nu_{kM}}$ now can be considered as function of the
phase $\delta,\rho,\sigma$,
the heavy majorana neutrinos masses and the complex angles $\widehat{\theta}_1,\widehat{\theta}_2,\widehat{\theta}_3$.
For simplicity, we assume $\widehat{\theta}_1=\widehat{\theta}_2=\widehat{\theta}_3\equiv \widehat{\theta}$.
In this work we consider the CP asymmetry due to the decays of the lightest
heavy Majorana $\nu_{1M}$. The detail will be presented in the subsections below.\\

 The baryon asymmetry and lepton asymmetry are given as
 \bea
 &&\eta_B=\fr{n_B-n_{\overline{B}}} s, \crn &&
 \eta_L=\fr{n_l-n_{\overline{l}}} s,
 \eea
 where $s$ is the entropy density.
 The lepton
 asymmetry can be transformed into a baryon asymmetry by non-perturbative
 B + L violating (sphaleron) processes \cite{sphaleron}, giving
 \be\eta_B = a(\eta_B-\eta_L)=\fr a {a-1}\eta_L,\ee
 where
 \be a=\fr{8 n_g + 4 n_H}{22 n_g + 13 n_H},\ee
 with $ n_H$ is the number of Higgs
 and
 $n_g$ is the number of fermion generations.
 We get \be\eta_B=- \fr{8 n_g + 4 n_H}{14 n_g + 9 n_H}\eta_L.\ee
 As the analysis in \cite{nH},
 taking $n_H=2$ and $n_g=3$, we get
 \be \eta_B=- \fr{8}{15}\eta_L\label{etaB}.\ee
Now let us calculate $\eta_L$ in thermal and non-thermal leptogenesis scenario.

\subsection{Thermal production}

In the thermal scenario, the heavy Majorana neutrinos are produced in a thermal bath.
At $T>m_{\nu_{1M}}$, the CP asymmetry generated by $\nu_{1M}$ decays can be washed out due to
inverse decays and scattering processes. That why the CP asymmetry is weighted by the washout efficiency.

For the channel $\nu_{kM}\rightarrow e_iH_5^+,\ \overline{e}_i H_5^-$ the CP asymmetry
depends on flavor because $L_i(e_i)=1$. However, since $L(N_i)=0, L(H')=-1$, the CP asymmetry
due to the decay $\nu_{kM}\rightarrow N_iH'^*,\ N_i H'$ is considered flavor independent.
The Boltzmann equations for the lepton asymmetry can be divided by two forms,
one is the equation for the flavored lepton asymmetry corresponding to $\varepsilon^{i(1)}_{\nu_{1M}}$, and another is treated by the conventional computation for
$\varepsilon^{(2)}_{\nu_{1M}}=\sum_i \varepsilon^{i(2)}_{\nu_{1M}}$.

The interference of the tree level with loop diagrams contained gauge propagator
is vanished in summation of all the indexes $i,l=1,2,3$ if the CP asymmetry has the
same weight for all flavors,
\be \sum_{i,l} \mathrm{Im}[ h^{\nu*} _{ik} h^\nu _{lk}]=\mathrm{Im}[\sum_{i,l}  h^{\nu*} _{ik} h^\nu _{lk}]=0.  \ee
Therefore, from Eq. (\ref{epsilon2})
\bea \varepsilon^{(2)}_{\nu_{1M}}=\sum_i \varepsilon^{i(2)}_{\nu_{1M}}&=&\fr{1}{8\pi C_0} \sum_{j} \sqrt{g_j}\left[1-(1+g_j)log[1+1/g_j]+ s_\beta^2(1-g_j)^{-1}\right]
 \mathrm{Im}[((h^{\nu\dag}h^\nu)_{1j})^2]
  \crn  &\simeq&-1.6\times 10^{-2} \sum_{j} \fr{\mathrm{Im}[((h^{\nu\dag}h^\nu)_{1j})^2]}{ \sqrt{g_j} (h^{\nu\dag}h^\nu)_{11}}.  \eea

Eq. (\ref{epsilon1}) can be reduced by taking  $g=0.65, s_W^2=0.231$ as
\be \varepsilon^{i(1)}_{\nu_{1M}}=\fr{-1.6\times 10^{-4}\sum_l \mathrm{Im}[h^{\nu *}_{i1}h^\nu_{l1}]-0.6\times 10^{-2} \sum_j \sqrt{g_j^{-1}}\mathrm{Im}[( h^{\nu\dag}h^\nu)_{1j}h^{\nu*} _{i1} h^\nu _{ij}]}{(h^{\nu\dag}h^\nu)_{11}}. \ee

In the thermal leptogenesis, the washout parameters  are defined as
\bea K_i&=&\fr{\Gamma(\nu_{1M}\rightarrow e_iH_5^+,\
\overline{e}_i H_5^- )}{H(T=m_{\nu_{1M}})}=\fr{s_\beta^2 h^{\nu*} _{i1} h^\nu _{i1} m_{\nu_{1M}}}{8 \pi} ,\crn
K&=&\fr{\Gamma(\nu_{1M})}{H(T=m_{\nu_{1M}})}=\fr{(2+s_\beta^2)(h^{\nu\dag}h^\nu)_{11} m_{\nu_{1M}}}{8 \pi}.
\eea
By varying $\delta, \sigma, \rho, \widehat{\theta}$
for $m_{\nu_{1M}} \sim 10^9$ GeV,
$m_{\nu_{2M}}\sim m_{\nu_{3M}} \sim 10^3 m_{\nu_{1M}} $ we figure out
$K\gg 1$ for all values of CP parameters $\delta, \sigma, \rho$, and the complex angle $\widehat{\theta} $.
The lepton asymmetry can be approximated as given in \cite{washout-para}.
In the strong washout regime, $K\gg 1$
\bea \eta_L^0 &\simeq & 0.3\fr{\varepsilon^{(2)}_{\nu_{1M}}}{g_*} \left(\fr {0.55}{K}\right)^{1.16}, \crn
\eta_L^i &\simeq & \fr{\varepsilon^{i(1)}_{\nu_{1M}}}{g_*}
\left(\fr{8.25}{|A_{ii}|K_i}+\left(\fr{|A_{ii}|K_i}{0.2}\right)^{1.16}\right)^{-1},
 \eea
 where $A_{11}=-151/179, \ A_{22}=A_{33}=-344/537$. 

\begin{figure}[!h]
\begin{center}
\includegraphics[width=8cm,height=8cm]{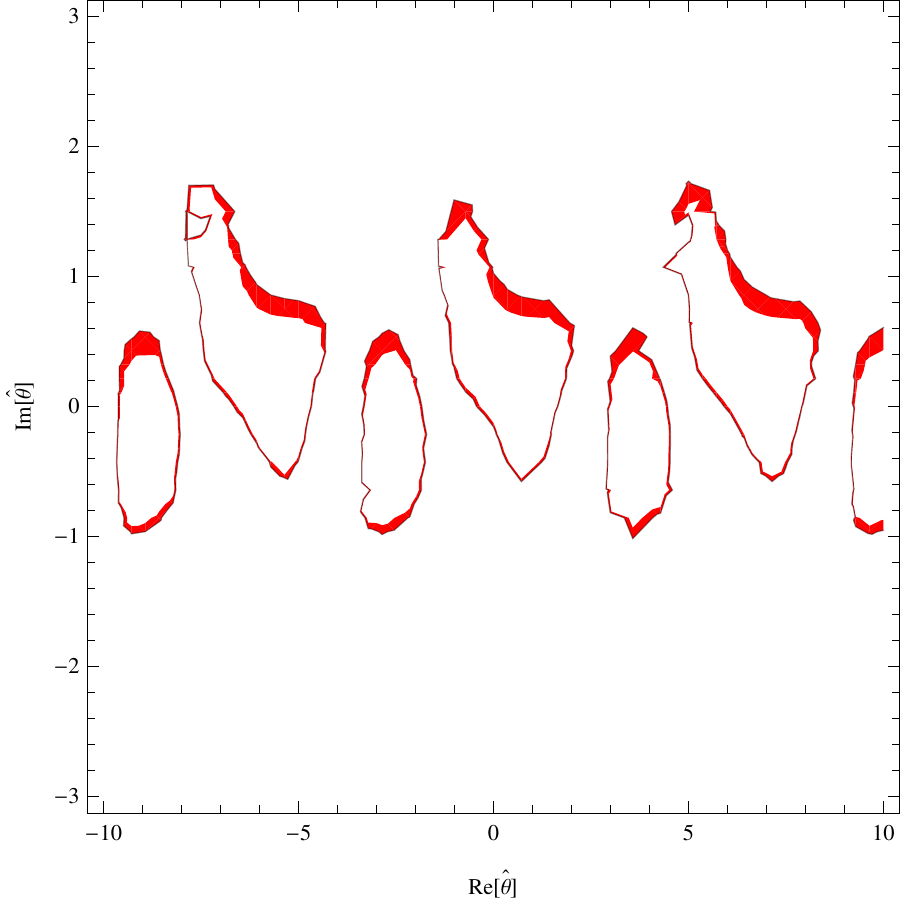}
\caption[]{\label{figetaBther} Contour plot of $\eta_B$ in the region $5\times10^{-11}<\eta_B< 10^{-10}$ on the plan of the complex angel $\widehat{\theta}$ for $\delta=4.3$ rad, $\sigma=-1.5$ rad, $ \rho=-1$ rad,
$m_{\nu_{2M}}=m_{\nu_{3M}}=10^3 m_{\nu_{1M}}$,  $  m_{\nu_{1M}}= 10^{9}$ GeV, $m_{\nu_1}=0.01$ eV.}
\end{center}
\end{figure}

\begin{figure}[!h]
\begin{center}
\includegraphics[width=8cm,height=6cm]{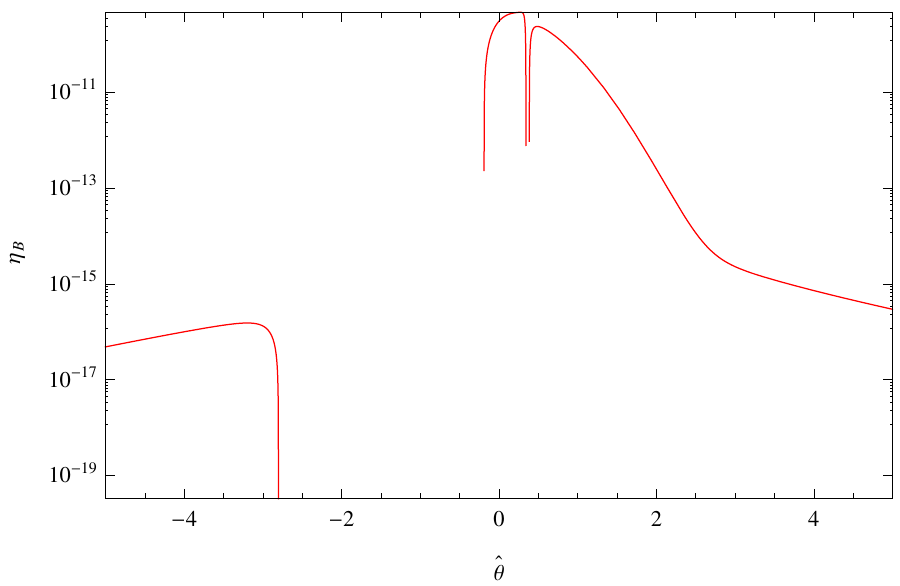}
\includegraphics[width=8cm,height=6cm]{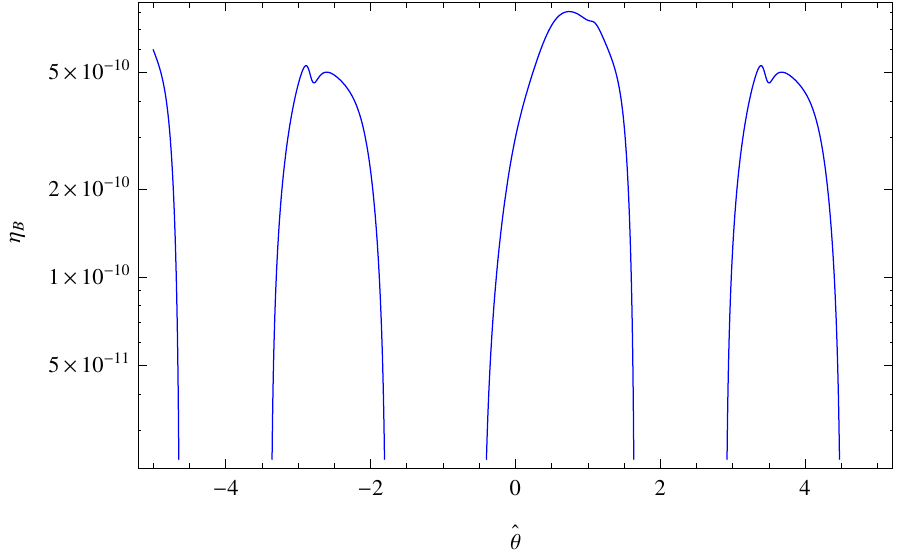}
\caption[]{\label{figImRether}  $\eta_B$ vs. pure imaginary $\widehat{\theta}$ (red) and pure real $\widehat{\theta}$ (blue)
for $\delta=4.3$ rad, $\sigma=-1.5$ rad, $ \rho=-1$ rad,
$m_{\nu_{2M}}=m_{\nu_{3M}}=10^3 m_{\nu_{1M}}$,  $  m_{\nu_{1M}}= 10^{9}$ GeV, $m_{\nu_1}=0.01$ eV.}
\end{center}
\end{figure}

\begin{figure}[!h]
\begin{center}
\includegraphics[width=8cm,height=6cm]{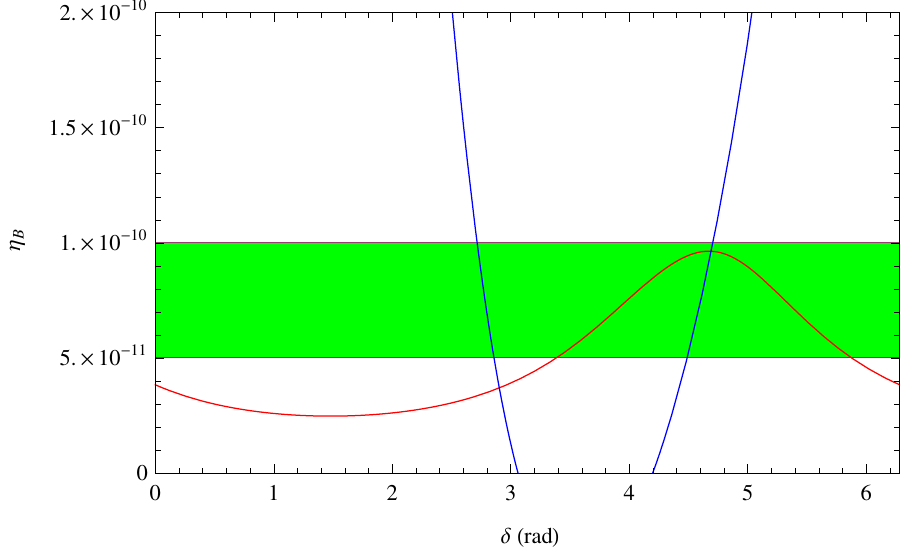}
\includegraphics[width=8cm,height=6cm]{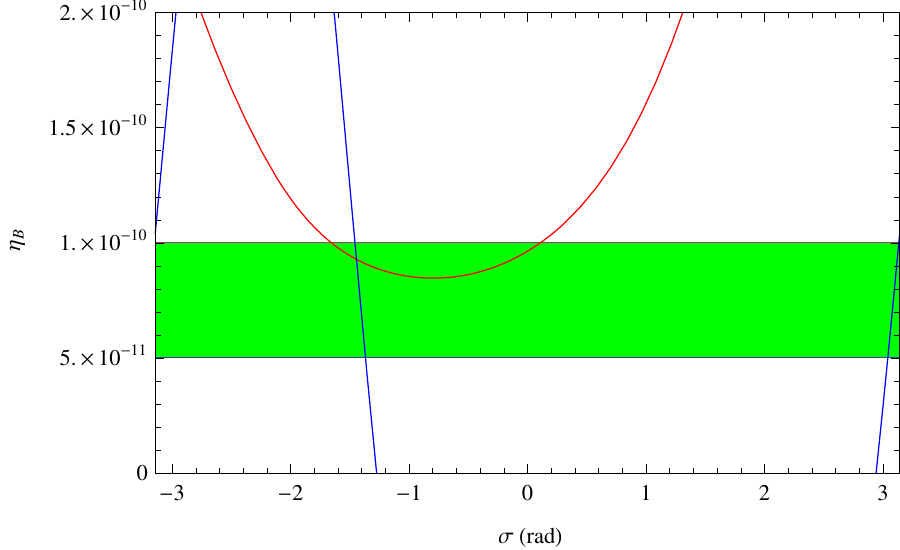}
\caption[]{\label{figCPther}  $\eta_B$ vs. $\delta$ (left) and  $\eta_B$ vs. $\sigma=\rho$
(right) for $\widehat{\theta}=0.87 I$ (red) and $\widehat{\theta}=-0.18 I$ (blue),
and other parameters given in (\ref{paraTher}).}
\end{center}
\end{figure}

The baryon asymmetry is related to the lepton asymmetry as
\be \eta_B=-\fr{8}{15}(\sum_{i=1,2,3} \eta_L^i+ \eta_L^0).\ee
From all expressions above, we see that $\eta_B$ depends on $\delta, \sigma, \rho$, and  $\widehat{\theta} $.
The baryon asymmetry $\eta_B$ in the region ($5\times10^{-11}, 10^{-10}$) on the plan
of the complex angel $\widehat{\theta}$ is shown in Fig. \ref{figetaBther} for
\bea &&\delta=4.3 \ \mathrm{rad},\hs \sigma=-1.5\ \mathrm{rad}, \hs \rho=-1\ \mathrm{rad},\crn
&& m_{\nu_{2M}}=m_{\nu_{3M}}=10^3 m_{\nu_{1M}},\hs  m_{\nu_{1M}}= 10^{9}\ \mathrm{GeV}, \hs m_{\nu_1}=0.01\ \mathrm{eV}.
\label{paraTher}
\eea
The red regions indicate that in order to satisfy $5\times10^{-11}<\eta_B<10^{-10}$,
we need to require $-1.01 < \mathrm{Im}[\widehat{\theta}] < 1.8 $ when varying Re$[\widehat{\theta}]$.
The limit of Im$[\widehat{\theta}]$ keeps the same if we extend the range of
Re$[\widehat{\theta}]$.
The $\eta_B$ is considered as function of pure imaginary $\widehat{\theta}$ (red) and pure real
$\widehat{\theta}$ (blue) as shown
in Fig. \ref{figImRether}. We see that $\eta_B$ changes a lot when varying  pure Im$[\widehat{\theta}]$ while it seems
to keep the same order when the pure Re$[\widehat{\theta}]$ alters.
The baryon asymmetry varies little as function of the CP phases presented in Fig. \ref{figCPther}.

If we study the case $m_{\nu_{1M}}=10^9 GeV, m_{\nu_{2M}}=m_{\nu_{3M}}= 10^5 m_{\nu_{1M}} $, the
constraint on the complex angle $\widehat{\theta}$ is stricter in order to satisfy the experimental
results on baryon asymmetry.

\subsection{Non-thermal production}

In the non-thermal scenario the reheating temperature can be lower than the lightest
heavy Majorana. The total CP asymmetry is the summation of all flavor CP asymmetry,
\be \varepsilon_{\nu_{kM}}=\sum_{i}(\varepsilon_{\nu_{kM}}^{i(1)}+\varepsilon_{\nu_{kM}}^{i(2)})= \fr{\sum_{j\neq k}B_j \mathrm{Im}[[(h^{\nu\dag} h^\nu )_{kj}]^2] }{  (h^{\nu\dag}h^\nu)_{kk} },\label{epsilonN}\ee
where
\bea B_j&=& \fr{1}{8 \pi (2+s_\beta^2)} \sqrt{g_j}\left[(s_\beta^4+1)(1-(1+g_j)log[1+1/g_j])+ s_\beta^2(s_\beta^2+1)(1-g_j)^{-1}\right] \crn &\simeq &-\fr{11}{160 \pi \sqrt{g_j}}. \eea
The lepton asymmetry is related with the CP asymmetry through
 \be\eta_L= \fr 3 2 \varepsilon_{\nu_{kM}} \times Br_k\times
 \fr{T_R}{m_\Phi}, \ee where  $Br_k$ denotes the branching ratio
 of the decay channel $\Phi \rightarrow \nu_{kM} \nu_{kM}$.

As analysis in the previous section, we assumed that $m_{\nu_{1M}}\ll m_\Phi< m_{\nu_{2M}}\sim m_{\nu_{3M}}$,
$m_\Phi<m_{Z^N}$ and $\Gamma (\Phi\rightarrow hh)\ll \Gamma (\Phi\rightarrow \nu_{1R}\nu_{1R})$
when $\la_{10;11;12}$ are negligibly small, therefore,
\be\eta_L\simeq \fr 3 2 \varepsilon_{\nu_{1M}}  \times
 \fr{T_R}{m_\Phi}. \ee
Combining Eqs. (\ref{TRnon}, \ref{etaB}, \ref{epsilonN}) with $u\sim v \sim 174\ \mathrm{GeV}$ we get
\be
\eta_B \simeq  0.4\times \fr{m_{\nu_{1M}}}{\sqrt2 <\Phi> }\times
\fr{\sum_{j=2,3}\fr{m_{\nu_{1M}}}{m_{\nu_{jM}}} \mathrm{Im}[[(h^{\nu\dag} h^\nu )_{1j}]^2] }
{ (h^{\nu\dag} h^\nu )_{11}},\ee
with notice that the formula $h^{\nu\dag} h^\nu $ given in Eq. (\ref{hnu1}).
Putting
\bea
&&\delta=4.3 \ \mathrm{rad},\hs \sigma=-1.5\ \mathrm{rad}, \hs \rho=-1\ \mathrm{rad},
\hs \widehat{\theta}=1.46 I, \crn
&&m_{\nu_{2M}}=m_{\nu_{3M}}=10^3 m_{\nu_{1M}},\hs  m_{\nu_{1M}}=2.34\times 10^{11}\ \mathrm{GeV}, \hs m_{\nu_1}=0.01\ \mathrm{eV},\crn
&& m_\Phi=2.67 \times 10^{13}\ \mathrm{GeV},\hs <\Phi>=23.6 m_P,\label{paraNon}\eea
 we get
\be
\eta_B \simeq  8.92  \times 10^{-11}.\ee
This value of baryon asymmetry is in agreement with \cite{etaBc},
 $\eta_B=(8.75 \pm 0.23) \times 10^{-11}$.

\begin{figure}[!h]
\begin{center}
\includegraphics[width=8cm,height=8cm]{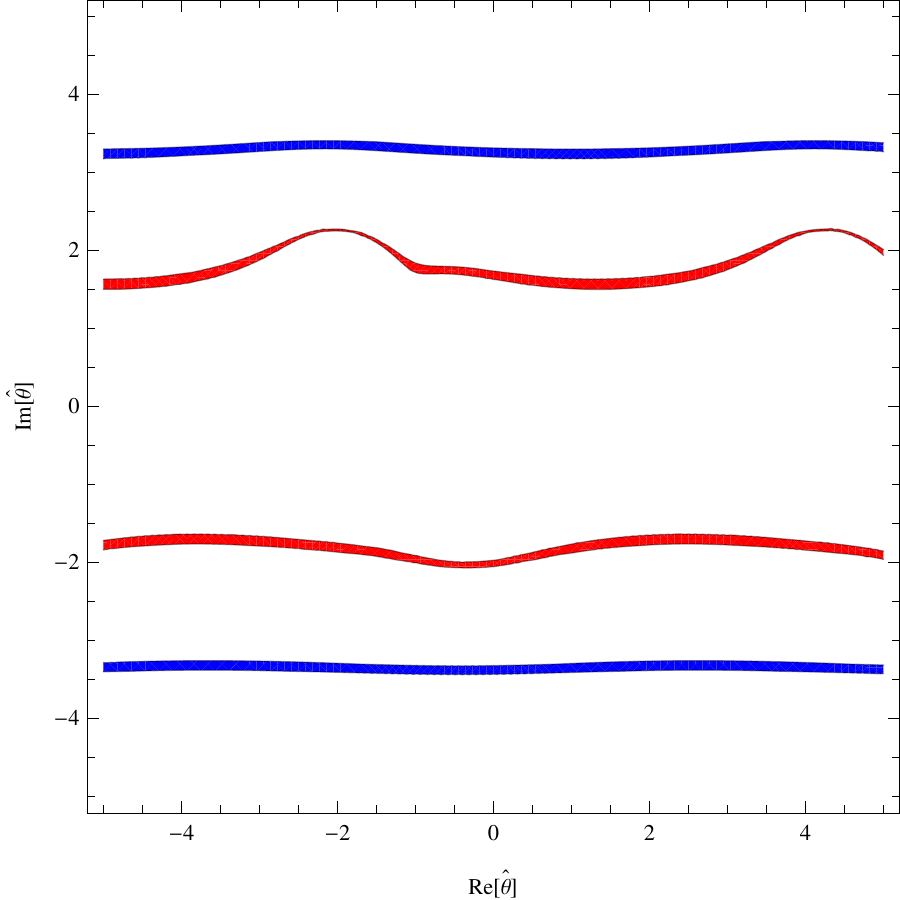}
\caption[]{\label{figetaBnon} Contour plot of $\eta_B$ in the region ($5\times10^{-11}<\eta_B< 10^{-10}$) on the plan of the complex angel $\widehat{\theta}$ for $m_{\nu_{1M}}= 10^{11}$ GeV (red)
and $m_{\nu_{1M}}= 10^{9}$ GeV (blue).}
\end{center}
\end{figure}

\begin{figure}[!h]
\begin{center}
\includegraphics[width=8cm,height=6cm]{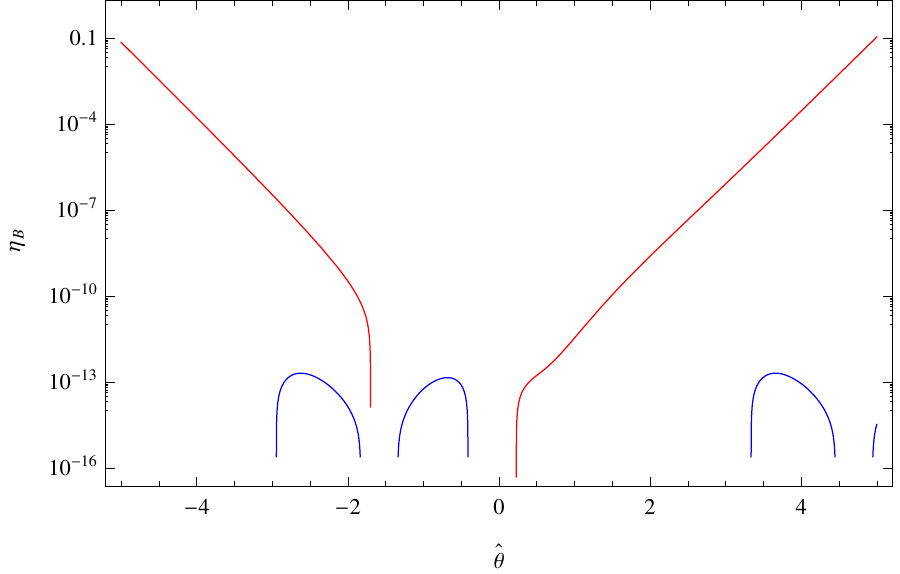}
\caption[]{\label{figImRe}  $\eta_B$ vs. pure imaginary (red) and pure real (blue) $\widehat{\theta}$.}
\end{center}
\end{figure}

Let us consider how $\eta_B$ depends on the complex angles and CP phases one by one.
Fig. \ref{figetaBnon} shows $\eta_B$ in the region ($5\times10^{-11}, 10^{-10}$) on the plan of the complex angel $\widehat{\theta}$  for
$m_{\nu_{2M}}=m_{\nu_{3M}}=10^3 m_{\nu_{1M}}$, $m_{\nu_{1M}}= 10^{11}$ GeV (red), and
$m_{\nu_{2M}}=m_{\nu_{3M}}=10^5 m_{\nu_{1M}}$, $m_{\nu_{1M}}= 10^{9}$ GeV (blue) and all
other parameters as given in (\ref{paraNon}). We see that in the red region
$-2.05 <\mathrm{Im}[\widehat{\theta}]< -1.68$ or  $1.49<\mathrm{Im}[\widehat{\theta}]< 2.28$  and in the blue region
$\mathrm{Im}[\widehat{\theta}]   \sim 3.3 $ or  $\mathrm{Im}[\widehat{\theta}]   \sim -3.4 $
 when varying Re$[\widehat{\theta}]$ even though if we extend the plot range for both axes. It means that it is free to choose
the value of Re$[\widehat{\theta}]$ but Im$[\widehat{\theta}]$ is quite a strict constraint.
$\eta_B$ depends strongly on Im$[\widehat{\theta}]$, while it changes lightly when varying Re$[\widehat{\theta}]$. This conclusion is more clearly in Fig. \ref{figImRe}, in which
$\eta_B$ is considered as a function of pure imaginary (red) and pure real  (blue) $\widehat{\theta}$ .

\begin{figure}[!h]
\begin{center}
\includegraphics[width=8cm,height=6cm]{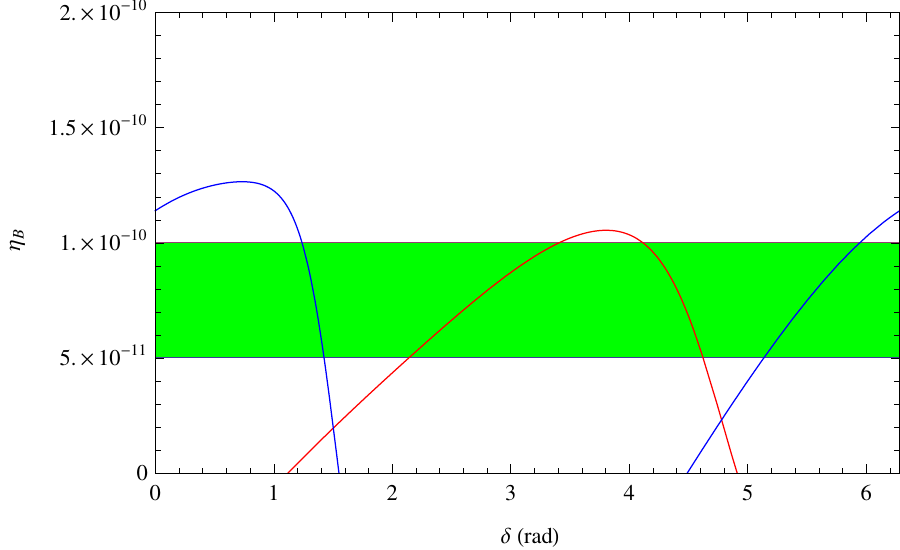}
\includegraphics[width=8cm,height=6cm]{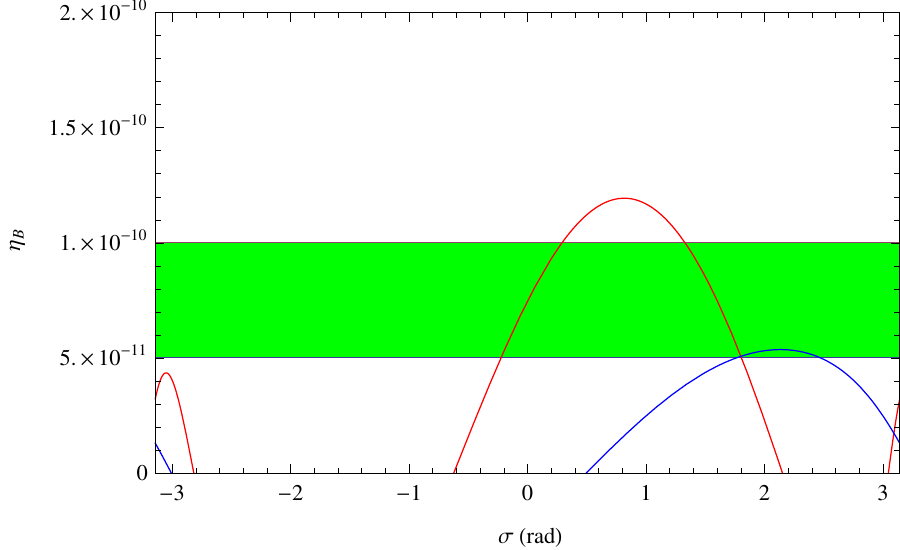}
\caption[]{\label{figCP} $\eta_B$ vs. $\delta$ (left) and  $\eta_B$ vs. $\sigma=\rho$
(right) for $\widehat{\theta}=1.46 I$ (red) and $\widehat{\theta}=-1.46 I$ (blue).}
\end{center}
\end{figure}

Fig. \ref{figCP} shows $\eta_B$ as a function of Dirac CP phase $\delta$ (left) and Majorana CP phase $\sigma=\rho$
(right) for $\widehat{\theta}=1.46 I$ (red) and $\widehat{\theta}=-1.46 I$ (blue), and the choice of other parameters given in (\ref{paraNon}). In brief, we see that $\eta_B$ does not
depend much on the CP phase but depend on the imaginary of the complex angle $\widehat{\theta}$.
This conclusion is the same as analysis in thermal scenario.

\section{\label{conc}Conclusions}

We have studied generation of inflation and leptogenesis in the 3-3-1-1 model
by considering the symmetry breaking of
the $U(1)_N$ gauge group at the GUT scale. The model contains two super heavy
particles with mass proportional to $\La$, the new gauge boson $Z^N$ embedded to $U(1)_N$
and the scalar Higgs boson $H_3\simeq S_4$. All other new massive particles get mass
in order of $\om$.
The singlet Higgs $\phi$ with $<\phi>$ at the GUT scale can play the role of inflaton. 
The quantum corrections to the potential of inflaton is taken into account, thus there appears
logarithm function of  inflaton, making the presently considered model's inflation different from chaotic one.
In this work, we have figured out the  parameter spaces appeared in the inflaton potential
matching the experiment on the spectrum index $n_s$, the tensor to scalar ratio $r$,
the running index $\alpha$ as well as the amplitude of the curvature perturbation $\Delta_\mathcal{R}^2$.
The  inflaton mass is obtained in an order of $10^{13}$ GeV.

After the inflation, the heavy Majorana can be produced in a thermal bath or
by decay of the inflaton. Depending on the Higgs couplings $\la_{10,11,12}$ in comparison with
the Yukawa couplings $h^{'\nu}_{ij}$, leptogenesis is considered in thermal
or non thermal scenario.
We have shown how the 3-3-1-1 model  generates
lepton asymmetry then converts into baryon asymmetry in both cases.
It is interesting that the model contains an extra channel contributing to the CP asymmetry.
The heavy Majorana
particles can decay into neutral neutrinos $N_i$ and neutral complex  Higgs $H'$
with the coupling
different by  factor $s_\beta$ from the original channel, $\nu_{kM}\rightarrow e_i^\pm H^\mp$.
In thermal leptogenesis, the CP asymmetry generated by the new channel is considered flavor independent,
while the ordinary channel is treated as flavor dependent due to the
different lepton number of $N_i$ and $e_i$. It leads the interference of
the tree level with loop diagrams appeared gauge propagator to contribute to the CP asymmetry for
the decay $\nu_{kM}\rightarrow e_i^\pm H^\mp$. This feature is new compared to
other leptogenesis models.

The thermal and non thermal leptogenesis
have been calculated in detail. In order to get non zero CP asymmetry
we need to consider the complex Yukwa coupling matrix $h^\nu$ by expressing it
in terms of the neutrino mass and mixing matrix and the orthogonal matrix $R$.
We have presented how
$\eta_B$ depends on the CP phases $\delta, \sigma, \rho$ and complex angle $\widehat{\theta}\equiv\widehat{\theta}_1=
\widehat{\theta}_2= \widehat{\theta}_3 $. The baryon asymmetry is not much
sensitive to
the value of CP phases or pure real $\widehat{\theta}$  but it alters a lot as a function of
pure imaginary $\widehat{\theta}$. This property is the same for both leptogenesis scenarios.
Thank to the orthogonal matrix $R$ and the complex angle  $\widehat{\theta}$, which makes the baryon symmetry
completely in agreement with the experiment for both cases.
One different thing of the two scenarios is that at any point of Re$[\widehat{\theta}]$
we always can find Im$[\widehat{\theta}]$ satisfying a fixed value of $\eta_B$ in non thermal case, but
there is restriction of choosing pair of (Im$[\widehat{\theta}]$, Re$[\widehat{\theta}]$)
in thermal scenario to match experiment on $\eta_B$. 
We know that the baryon asymmetry depends much on Im$[\widehat{\theta}]$ and
it is easy to see that from the Fig. \ref{figImRether},
there is an upper limit on the baryon asymmetry if we consider  $\eta_B$ as a function of
Im$[\widehat{\theta}]$ in thermal scenario because of the effect of washout efficiency.
However, there is no upper bound for $\eta_B (\mathrm{Im}[\widehat{\theta}] )$ in non thermal case,
see Fig. \ref{figImRe}.
By considering non thermal leptogenesis, the reheating temperature $T_R$ can be reduced much
lower than the lightest heavy Majorana mass.
In brief, the 3-3-1-1 model at the GUT scale successfully explains the baryon
asymmetry of the universe by studying both thermal and non thermal leptogenesis
mechanisms.

\section*{Acknowledgments}

This research is funded by Vietnam National Foundation for Science and Technology Development
(NAFOSTED) under grant number 103.01-2014.69, and by the National Research Foundation of Korea (NRF)
grant funded by Korea government of the Ministry of Education, Science and
Technology (MEST) (No. 2011-0017430) and (No. 2011-0020333).

\end{document}